\documentclass[12pt,draftcls,onecolumn]{IEEEtran}

\newcounter{mytempeqncnt}
\usepackage{algorithm2e}
\usepackage{cite}
\usepackage{amsmath} 
\usepackage{amssymb} 
\usepackage{dsfont}
\usepackage[mathcal]{euscript}
\usepackage{empheq}
\usepackage{graphicx}
\usepackage{graphics}
\usepackage{bm}
\usepackage{psfrag}
\usepackage{mathrsfs}
\usepackage{color}
\usepackage{cancel}
\input{epsf}

\renewcommand{\P}{\mathbb{P}}
\newcommand{\E}{\mathbb{E}}
\newcommand{\beq}{\begin{equation}}
\newcommand{\eeq}{\end{equation}}
\newcommand{\beqa}{\begin{eqnarray}}
\newcommand{\eeqa}{\end{eqnarray}}
\newcommand{\dfz}{\triangleq}

\newcommand{\VAR}{\textnormal{VAR}}

\begin{document}
%\title{Who's Who in the Network: \\Botnet Identification under Random DDoS Attacks} %with Increasing Emulation Dictionary}
\title{DDoS Attacks with Randomized Traffic Innovation: Botnet Identification Challenges and Strategies}

\author{Vincenzo~Matta, Mario~Di Mauro, and Maurizio~Longo
\thanks{A short and limited version of this work appears in the conference publication~\cite{MDML_EUSIPCO2016}.}
\thanks{The authors are with DIEM, University of Salerno,
via Giovanni Paolo II, I-84084, Fisciano (SA), Italy (e-mail: \{vmatta, mdimauro, longo\}@unisa.it).
}
}

\maketitle

%%%%%%%%%%%%%%%%%%%%%%%%%%%%%%%%%%%%%%%%%%%%%%%
\begin{abstract}
Distributed Denial-of-Service (DDoS) attacks are usually launched through the {\em botnet}, an ``army'' of compromised nodes hidden in the network. 
Inferential tools for DDoS mitigation should accordingly enable an early and reliable discrimination of the normal users from the compromised ones. 
Unfortunately, the recent emergence of attacks performed at the application layer has multiplied the number of possibilities that a botnet can exploit to conceal its malicious activities. 
New challenges arise, which cannot be addressed by simply borrowing the tools that have been successfully applied so far to earlier DDoS paradigms. 
In this work, we offer basically three contributions: 
$i)$ we introduce an abstract model for the aforementioned class of attacks, where the botnet emulates normal traffic by continually learning admissible patterns from the environment; 
$ii)$ we devise an inference algorithm that is shown to provide a consistent (i.e., converging to the true solution as time elapses) estimate of the botnet possibly hidden in the network; 
and $iii)$ we verify the validity of the proposed inferential strategy over {\em real} network traces.
\end{abstract}

\vspace*{-5pt}

\vspace*{-5pt}
\section{Introduction and Motivation}
Cyber-security ranks among the biggest challenges of modern times. 
Whether we are talking of phishing, website sabotages, or even of terrorist attacks, protecting our digital lives is an issue of paramount importance. Networks, and especially the Internet, became the natural attackers' habitat to hide a broad variety of threats. 
For instance, a dangerous attack to a powerful target site (e.g., a big e-commerce portal) is often launched through a series of apparently innocuous attacks to some powerless, but most vulnerable, sites (e.g., some personal computers). 

One of the most popular threats is the Denial-of-Service (DoS) attack, which can be broadly categorized as a {\em volumetric} attack, where the target destination is overwhelmed  by a huge number of requests, eventually leading to the impossibility of serving any of the users. 
In particular, with a {\em Distributed} DoS (DDoS) attack, such a huge number of requests is produced in parallel by a net of robots (the {\em botnet}).
According to one of the classical DDoS representations, a relatively large ensemble of machines (the {\em bots} or {\em zombie} ``army''), acts cooperatively under the supervision of one or more coordinators (the {\em botmasters}).
The bots may be either themselves malicious users acting consciously, or they may be legitimate users that have been preliminarily infected, (e.g., by warms and/or Trojans). 

The existence itself of an anomalous request rate is essentially uncovered, and, hence, its detection is not a big deal. 
The main challenge is instead ascertaining whether the anomaly is caused by a DDoS attack, and, if so, performing a correct/early identification of the botnet hidden in the network. These operations are crucial to achieve successful DDoS mitigation, since discriminating legitimate from malicious users would allow the destination to ban the latter, without denying the service to the former. 
Providing inference solutions to botnet discovery and identification is the main subject of this work.

\vspace*{-10pt}
\subsection{Related Work}
The literature about DDoS attacks is rich, and we refer the Reader to the survey in~\cite{IEEEComSurveyTut} as a useful entry-point. The earliest DoS paradigms (see, e.g., TCP SYN flooding), relied on specific protocols' vulnerabilities, and were characterized by the repetition of one (or few) requests with a huge rate.
In this situation, the single source of the attack can be identified by simply computing its unusually large request rate. 

The {\em distributed} variants of such attacks exploit basically the same kind of vulnerabilities and repetition schemes, but for the fact that the {\em large} request rate is now obtained by aggregating {\em many small} individual bot rates. 
%the requests reach the target destination through a large number of bots. 
%In this way, while the overall request rate produced by the botnet is huge, 
%in order to avoid an identification based on the traffic volume at a single-user level. 
This notwithstanding, in such attacks, the bots can be still identified at a single-user level. 
Indeed, normal traffic patterns are typically characterized by a certain degree of innovation, while the repetition scheme implicitly emphasizes the bot character. 
In fact, several useful inferential strategies have been proposed for such kind of DDoS attacks, see~\cite{IEEEComSurveyTut} for a comparative summary. 

Recently, the new class of {\em application-layer} DDoS attacks is emerging as one of the most powerful threats~\cite{sucuri,riorey,incapsula,ferretti}. 
In such attacks, the malicious traffic patterns are disguised as normal ones by leveraging the many possibilities offered at the application layer (for instance, when surfing through a website, more and more web-pages are likely to be explored as time elapses). 
By assigning a sufficient degree of variability to each individual bot's pattern, identification strategies based on single-user inspection become harmless. 
% and, therefore, the repeatability issue is significantly reduced. 
Building on such new possibilities, in this work we shall introduce a formal model for DDoS attacks where the botnet gets at its disposal a certain {\em emulation dictionary} to build the traffic patterns. %The assumption is made that the botnet is able to continually enrich its knowledge by increasing, as time elapses, the number of messages forming the emulation dictionary.

A number of intriguing questions arise. 
Despite the strong power given to the attacker, is it still possible to consistently unveiling the presence of a botnet? 
If so, which are the pertinent inferential strategies, and which the performance limits? 
%Given the lack of information about the underlying statistical models, are such strategies universal and non-parametric?
%Are they computationally affordable? 
Which is the fundamental trade-off between the botnet learning ability and the inference performance?

\subsection{Relevant Inferential Tools and Methods}
The inferential strategies available in the literature are not conceived to manage the class of DDoS attacks with increasing emulation dictionary~\cite{IEEEComSurveyTut}. 
While in principle it is possible to generalize and take inspiration from some of these strategies, plug-in solutions to our problem are currently unavailable. 
Therefore, new inferential solutions must be conceived. 

Classical parametric statistical methods (e.g., maximum likelihood, Neyman-Pearson tests) typically offer a high degree of tractability, analytical results and performance guarantees, but they are suited to those situations where a detailed knowledge of the models is available~\cite{MaranoMattaTongByz,MaranoMattaWillettPhysLayer,BarniTIFS2013,VarshneyByz}, a condition that is far from being  verified in our setting. 
As a result, the aforementioned benefits are often paid in the coin of scarce robustness, sensible performance loss, low degree of versatility and adaptation. 

In contrast, fully data-driven techniques (e.g., machine learning) do not require a detailed knowledge of the underlying models, and provide a high degree of versatility, with these advantages being often paid in the coin of lack of analytical results and performance guarantees, hard physical interpretation of the metrics, heavy algorithm-tuning when parameters change. 

In order to partly circumvent the limitations of both approaches, as well as to retain some advantages thereof, in the present work we follow some emerging trends in signal processing for network cyber-security applications, which lie somehow in-between parametric and fully data-driven techniques.
As notable examples, we mention: sparsity-aware algorithms for unveiling traffic volume anomalies~\cite{Giannakis1,Giannakis2,Giannakis3}; universal algorithms for tracing information flows across the network~\cite{Venk-He-Tong-IT, He-Tong-Forens,He-Tong-IT,He-Agaskar-Tong-SP, Kim-Tong-SP2012, Marano-Matta-He-Tong-IT}; hypothesis testing in the presence of adversaries that can corrupt the data, when the statistical hypotheses are specified only through training data~\cite{BarniIT2014}.
Inspired by a common underlying philosophy, such works suggest to pursue the following principled approach: $i)$ focus on minimal-and-realistic physical assumptions; $ii)$ envisage physically-meaningful descriptive indicators arising from the modeling assumptions; $iii)$ devise consequently an inference strategy.

%%%
The DDoS class considered in this work builds upon and generalizes some dangerous threats that have been recently documented in the literature. 
To the best of our knowledge, this is the first attempt to provide a systematic analysis and to devise suitable countermeasures for such kind of attacks. 
As a future research step, an interesting extension would be optimizing (from the attacker's viewpoint, i.e., playing the bad guy role) the kind of DDoS attack.
This approach would naturally lead to an {\em adversarial} perspective where the DDoS and the botnet-identification strategies
should be optimized jointly, by looking for equilibrium solutions aimed at managing the attacker's and defender's conflicting requirements~\cite{BarniICASSP2013}.

\subsection{Main Result}
This work deals with the design and analysis of inference strategies aimed at identifying a botnet in the context of {\em distributed} denial-of-service attacks. 
In our setting: $i)$ the network analyst collects traffic patterns from across the network, and has access to the message content; 
$ii)$ the meaning of the messages produced by an individual user provides no special information about its nature, legitimate, or malicious; 
and $iii)$ no specific assumptions are made about the characterization of the traffic patterns of a normal user. 
%In particular, our results do not rely on any statistical model for the normal traffic patterns. 
In this respect, the inference strategies proposed in this work are non-parametric.    

Starting from the attacks documented in the literature, we introduce a formal model for randomized DDoS attacks with increasing emulation dictionary, which is defined by the following main features: 
$i)$ the botnet emulates the normal traffic patterns by gleaning admissible messages from an emulation dictionary; 
and $ii)$ the botnet is given the strong power of learning an emulation dictionary that becomes richer and richer as time elapses, so as to guarantee a sufficient variability across messages. 
In order to quantify the botnet {\em learning ability}, in this work we introduce the Emulation Dictionary Rate (EDR), namely, the increase of dictionary cardinality per unit time. 

Notably, the considered class of DDoS attacks is more general and powerful than many attacks documented in the literature.
The assumption of such great power in the attacker's hands might perhaps look overly pessimistic. 
At the same time, a worst-case analysis is perfectly suited to security applications, and allows getting important insights as regards the botnet identifiability under challenging operational conditions.
%Otherwise stated, the conducted analysis reveals how a network would be able to react to a powerful randomized DDoS with increasing emulation dictionary.

%The novelty of the class of DDoS attacks has two fundamental consequences. 
%First, as far as we can tell, documented cases of randomized DDoS with increasing emulation dictionary are as yet unavailable. Accordingly, when testing the botnet identification algorithms, the normal-users' activities are taken from {\em real-network} traces, while the application-layer DDoS are generated following the prescription of belonging to the introduced novel class.

The fundamental descriptive indicator employed in this work to ascertain the nature of network users is the Message Innovation Rate (MIR), namely, the number of distinct messages per unit time, transmitted by a given group of users.
The relevance of the MIR for botnet identification purposes arises since, in view of the coordination in the DDoS attack, the users belonging to a botnet are expected to exhibit a smaller degree of innovation than normal users, which act by their own nature independently one each other.

Our first contribution determines the MIR for a botnet $\mathcal{B}$, with either deterministic or Poisson transmission scheduling. Denoting by $\lambda_{\mathcal{B}}$ the transmission rate corresponding to the overall transmission activity in $\mathcal{B}$, and by $\alpha$ the EDR, we show that the MIR converges in probability to the following innovation rate (Theorem~$1$):
\beq
\boxed{
\mathscr{R}(\alpha,\lambda_{\mathcal{B}})=\frac{\alpha\,\lambda_{\mathcal{B}}}{\alpha+\lambda_{\mathcal{B}}}
}
\eeq
Our second contribution consists of devising an algorithm that, under a suitable Botnet Identification Condition (BIC), guarantees that the botnet hidden in the network is correctly identified as time elapses (Theorem~$2$). 

Finally, as a third contribution, all of the aforementioned theoretical results are tested and validated over {\em real} network traces; the experimental outcomes are definitely encouraging.

\noindent {\bf Notation}. $\P[\cdot]$ and $\E[\cdot]$ denote the probability and the expectation operators, respectively.  
Given an ensemble of random variables $X_t$ (with either continuous or discrete index $t$), the notation $X_t\stackrel{\textnormal{p}}{\longrightarrow}X$ means that $X_t$ converges in probability to $X$ as $t\rightarrow\infty$~\cite{shao}.

\section{Network Activity Indicators}
We start by introducing the basic quantities that will be used to describe the network activity. 
The first quantity relates to the {\em transmission} activity of the network users. 
Each user employs a certain scheduling, which is identified by the transmission epochs of its own messages. 
More in general, for any given subnet $\mathcal{S}$ of the network, we can define the {\em aggregate} pattern that comprises all (ordered) transmission epochs of the users belonging to $\mathcal{S}$, formally: $T_{\mathcal{S}}(1),T_{\mathcal{S}}(2),\dots$, where $T_{\mathcal{S}}(i)$ is the $i$-th (random) transmission epoch of users belonging to $\mathcal{S}$. 
Likewise, the pattern of an individual user $u$ becomes: $T_u(1),T_u(2),\dots$, where, with a slight abuse of notation (which will be used throughout the work), we have written $u$ in lieu of $\{u\}$.
The total number of transmissions occurred in $\mathcal{S}$, up to a given (deterministic) time $t$ is denoted by 
$N_{\mathcal{S}}(t)\dfz |\{i:T_{\mathcal{S}}(i)\leq t\} |$. 

As an indicator of the {\em transmission} activity, we introduce the {\em empirical} transmission rate at time $t$, namely,
\beq
\boxed{
\hat\lambda_{\mathcal{S}}(t)\dfz\frac{N_{\mathcal{S}}(t)}{t}
}
\label{eq:lambdaest}
\eeq
Whenever a limiting rate (as $t$ goes to infinity) is meaningfully defined, it will be denoted by $\lambda_{\mathcal{S}}$, which will be simply referred to as the transmission rate of subnet $\mathcal{S}$. 

Two examples of transmission schedulings which are relevant for our DDoS application, and which admit a limiting rate, are the synchronous, constant-rate transmission scheduling, and the independent Poisson scheduling.
In the former case, all users transmit synchronously, and the (constant) interval between two transmissions has duration $1/\lambda$. The empirical transmission rate clearly obeys:
$\hat\lambda_{\mathcal{S}}(t)\rightarrow\lambda\,|\mathcal{S}|$ as $t\rightarrow\infty$.
In the latter case, the transmission pattern of user $u$ is a Poisson process with rate $\lambda_u$, and the processes are mutually independent. Since the aggregate of independent Poisson processes is still a Poisson process, as a straightforward application of the (weak) law of large numbers, we have~\cite{ross-book}: $\hat\lambda_{\mathcal{S}}(t)\stackrel{\textnormal{p}}{\longrightarrow}\sum_{u\in\mathcal{S}}\lambda_u$.

As a second indicator of the network activity, we define a quantity that relates to the {\em content} of the messages sent by network users. We are interested in the {\em new} messages that are incrementally produced by the users during their activities, namely, in a Message Innovation Rate (MIR).
In order to obtain a formal definition of the MIR, let $\mathscr{D}_{\mathcal{S}}(t)$ denote the empirical dictionary composed by the {\em distinct} messages sent, up to time $t$, by users within $\mathcal{S}$. 
For the sake of clarity, we remark that, if the same message is sent, e.g., twice, from users belonging to $\mathcal{S}$, it appears only once in the dictionary $\mathscr{D}_{\mathcal{S}}(t)$.  
The {\em empirical} Message Innovation Rate (MIR) is:
\beq
\boxed{
\hat\rho_{\mathcal{S}}(t)\dfz\frac{|\mathscr{D}_{\mathcal{S}}(t)|}{t}
}
\label{eq:rhoest}
\eeq
In particular, if $\hat\rho_{\mathcal{S}}(t)\stackrel{\textnormal{p}}{\longrightarrow}\rho_{\mathcal{S}}$, the limiting value $\rho_{\mathcal{S}}$ will be simply referred to as the MIR of subnet $\mathcal{S}$.

\section{Randomized DDoS with Emulation Dictionary}
\label{sec:bibiebibotnet}
A botnet $\mathcal{B}_{\textnormal{tot}}$, composed by $B_{\textnormal{tot}}$ malicious nodes, sends messages to the destination under attack in order to saturate its resources. 
%As already said, we shall always imply that the mere content of a message does not reveal any particular information about the user's nature (legitimate or malicious).
%, and, hence, the traffic analyst is not able to discriminate legitimate from malicious users by examining only the individual users'  packets. 
%We note in passing that such assumption allows the botnet to track the dynamically changing contents in the attacked site (otherwise a bot would send inadmissible message, becoming easily identifiable), and 
%For the challenging case that the site under attack has a dynamical structure, the dictionary cannot be constructed beforehand (or, at least, not too in advance), 
The botnet mimics normal patterns by picking messages from an emulation dictionary, which is learned {\em continually} (i.e., its cardinality increases with time), in order to ensure that a reasonable innovation rate can be sustained. 
Such a dictionary construction can occur in many different ways. For instance, by means of one or more powerful botmasters, the botnet might be able to perform an on-line monitoring of normal activities from across the network. From such a monitoring, sequences of messages corresponding to normal patterns of activity are collected, allowing the construction of a dictionary of admissible messages.

Let $\mathscr{E}(t)$ be the (common) dictionary available at time $t$ to all botnet members. 
%We assume that the dictionary is initially not empty, namely, that $d_0\dfz|\mathscr{D}^\star_t|\geq 1$, and that the dictionary has a (linear) growth rate $\alpha$, namely,
We assume that the number of messages available for emulation grows, asymptotically, in a linear fashion. 
Therefore, it makes sense to introduce the Emulation Dictionary Rate (EDR) as:
\beq
\boxed{
\alpha\dfz \lim_{t\rightarrow\infty} 
\frac{|\mathscr{E}(t)|}{t}
}
\label{eq:EDRdef0}
\eeq 
%\beq
%\boxed{
%|\mathscr{D}^\star_t|=\lfloor d_0+\alpha t\rfloor
%}
%\eeq 
Given the emulation dictionary, the botnet has clearly many ways to build the traffic patterns. 
At one extreme, the botmaster disseminates $B_{\textnormal{tot}}$ {\em disjoint} (say, equal-sized) portions of $\mathscr{E}(t)$ through the botnet. Then, each bot builds its traffic pattern by scanning, in a sequential fashion, its portion of the emulation dictionary. Such a scheme would clearly maximize the independence among the bots. 
With this policy, the problem would become equivalent to the case that each bot owns a distinct emulation dictionary with EDR equal to $\alpha/B_{\textnormal{tot}}$.
However, since $B_{\textnormal{tot}}$ must be large, it is unrealistic to assume that a botmaster can learn so many patterns to build $B_{\textnormal{tot}}$ distinct dictionaries that are in turn so rich to guarantee a credible emulation.  
Therefore, in the case of disjoint dictionaries, the number of distinct messages available to a single bot would be typically small, implying a suspiciously high degree of replication, which would make the bots easily identifiable by single-user inspection.

At the other extreme, each bot might simply use {\em all} messages contained in $\mathscr{E}(t)$. 
Clearly, such scheme maximizes the innovation of each individual bot, but also maximizes the dependence inside the botnet. 
By inspection of the messages sequentially sent by two or more bots, a traffic analyst would recognize an anomalous behavior.

We hence assume that the attacker has devised some intermediate strategy to circumvent the aforementioned issues. 
We introduce a class of {\em randomized} DDoS attacks, where a bot that intends to transmit at time $t$ picks a message from the available emulation dictionary $\mathscr{E}(t)$, and sends such a message to the destination. The message is chosen uniformly at random, so that the probability of a particular message is simply $1/|\mathscr{E}(t)|$. 

The corresponding evolution of the empirical dictionaries, for any subnet $\mathcal{B}$ of $\mathcal{B}_{\textnormal{tot}}$, is easily obtained as follows. 
Given the empirical dictionary $\mathscr{D}_{\mathcal{B}}(t)$, the empirical dictionary $\mathscr{D}_{\mathcal{B}}(t+\tau)$ is obtained by adding the {\em distinct} messages not contained in $\mathscr{D}_{\mathcal{B}}(t)$, which have been selected during the interval $\tau$ by the bots belonging to $\mathcal{B}$.

%According to the above description, given the number of distinct messages $M_{n-1}$ at time $n-1$, during the $n$-th slot, the number of distinct messages (possibly) increases by the number of {\em distinct} messages picked by the bots, which in turn do {\em not} belong to the current empirical dictionary $\tilde{\mathscr{D}}_{n-1}$. 
We stress that the scheme examined here is not the only one. 
Other possible attacks include: purely volumetric DDoS; strategies using disjoint dictionaries; hybrid strategies using groups of disjoint dictionaries disseminated through the network. 
While in this work we focus on a specific class of DDoS attacks, our treatment (as will be clear from the forthcoming analysis) is sufficiently flexible to accommodate extensions to many interesting scenarios.

\subsection{Characterization of the Botnet Message Innovation Rate}
Let us preliminarily introduce the following function:
\beq
\boxed{
\mathscr{R}(\alpha,\lambda)\dfz\frac{\alpha\,\lambda}{\alpha+\lambda}
}
\label{eq:Rfundef}
\eeq
Our first result provides a closed-form expression for the MIR of a botnet. 

\noindent
{\bf \textsc{Theorem 1} (Botnet MIR).}
{\em 
Consider a botnet $\mathcal{B}_{\textnormal{tot}}$ launching a DDoS attack, where the node transmission policies are either synchronous with constant transmission rate, or independent Poisson processes, with rates $\lambda_u$, for $u\in\mathcal{B}_{\textnormal{tot}}$. 
Consider a subset of the botnet $\mathcal{B}\subseteq \mathcal{B}_{\textnormal{tot}}$.
Let $\mathscr{E}(t)$ be the emulation dictionary available to the botnet, with emulation dictionary rate $\alpha$, and let $\mathscr{D}_{\mathcal{B}}(t)$ be the empirical dictionary of the subnet $\mathcal{B}$ at time $t$. 
Then, the message innovation rate of $\mathcal{B}$ is:
\beq
\boxed{
\frac{|\mathscr{D}_{\mathcal{B}}(t)|}{t}\stackrel{\textnormal{p}}{\longrightarrow}
\rho_{\mathcal{B}}=\mathscr{R}(\alpha,\lambda_{\mathcal{B}})
}
\label{eq:theor2}
\eeq
where $\lambda_{\mathcal{B}}=\sum_{u\in\mathcal{B}}\lambda_u$ is the aggregate transmission rate of the considered botnet subset.
}

{\em Proof}. See Appendix~\ref{app:A}.~$\hfill\blacksquare$

\noindent
\textsc{Remark I.}
From~(\ref{eq:Rfundef}) and~(\ref{eq:theor2}) we see that increasing the EDR $\alpha$ and/or the transmission rate $\lambda$ corresponds to increasing the MIR. 
Besides, the MIR is always smaller\footnote{For $x>0$ and $y>0$, one has $x/(x+y)\leq 1$.} than $\min(\alpha,\lambda)$, which makes sense, since the number of new messages can exceed neither the number of messages in the emulation dictionary ($\mathscr{R}(\alpha,\lambda)\leq \alpha)$, nor the overall number of transmitted messages ($\mathscr{R}(\alpha,\lambda)\leq \lambda$). 
Notably, the quantity $\min(\alpha,\lambda)$ is the MIR corresponding to a practical scheme where the patterns are obtained by taking sequentially (in a deterministic way) the messages of the emulation dictionary. 
With such a scheme, if $\alpha>\lambda$, a new message can be always found in $\mathscr{E}(t)$, and the maximum rate of distinct messages is $\lambda$. 
Likewise, if $\lambda>\alpha$, all messages in $\mathscr{E}(t)$ can be selected, along with some unavoidable repetitions, and the maximum rate of distinct messages is $\alpha$. 

\noindent
\textsc{Remark II.} 
As $\alpha$ goes to infinity, the MIR converges to $\lambda$. In fact, as the number of messages in the emulation dictionary goes to infinity, each transmission would correspond with high probability to a new message, and the MIR will eventually reach the maximum allowable value $\lambda$.
Likewise, as $\lambda$ goes to infinity, we see that the MIR converges to $\alpha$. In fact, as the number of sent messages goes to infinity, the emulation dictionary is completely spanned, and the MIR will eventually saturate to its maximum allowable value $\alpha$.

\noindent
\textsc{Remark III.}
The MIR is symmetric in $\alpha$ and $\lambda$, implying that both quantities, even if they have a completely different practical meaning, play the same role as regards their effect on the MIR. 
In particular, we can write $\mathscr{R}(\alpha,\lambda)=(1/\alpha + 1/\lambda)^{-1}$, which reveals that the rate $\mathscr{R}(\alpha,\lambda)$ can be represented as the inverse of a time interval given by the sum of the average time between two messages available in the emulation dictionary, $1/\alpha$, and the average time between two transmissions, $1/\lambda$.

\noindent
\textsc{Remark IV.}
For strictly positive $\alpha$ and $\lambda$ we have:
\beq
\boxed{
\mathscr{R}(\alpha,\lambda_1)+\mathscr{R}(\alpha,\lambda_2)
>
\mathscr{R}(\alpha,\lambda_1+\lambda_2)
}
\label{eq:magicbound}
\eeq
The latter inequality can be straightforwardly checked by exploiting the definition of $\mathscr{R}(\alpha,\lambda)$ in~(\ref{eq:Rfundef}). More interestingly, such inequality can be explained in the light of the physical interpretation of Theorem~$1$. 
In fact, the LHS in~(\ref{eq:magicbound}) corresponds to the MIR of a botnet made of two subnets: $i)$ featuring transmission rates $\lambda_1$ and $\lambda_2$, respectively, and $ii)$ picking messages from two {\em disjoint} dictionaries, each one with EDR equal to $\alpha$. 
In contrast, the RHS corresponds to the MIR of a botnet made of two subnets, still featuring transmission rates $\lambda_1$ and $\lambda_2$, but picking messages from a {\em common} dictionary with EDR $\alpha$. 
Hence, the lower bound follows. 

\noindent
\textsc{Remark V.}
Our focus is on {\em genuinely-distributed} DoS attacks where the number of bots is large, {\em and} the transmission rate of each bot is not anomalous. 
Let us now consider a different DDoS strategy. 
Assuming for simplicity that all bots have unitary transmission rates, the MIR of user $u$, and the MIR of the whole botnet will be, respectively,
\beq
\rho_u=\frac{\alpha}{\alpha+B_{\textnormal{tot}}}, \quad
\rho_{\mathcal{B}_{\textnormal{tot}}}=\sum_{u\in\mathcal{B}_{\textnormal{tot}}} \rho_u=\frac{\alpha\,B_{\textnormal{tot}}}{\alpha+B_{\textnormal{tot}}},
\label{eq:singledisjoint}
\eeq
where the first relationship follows from Theorem~1, while the second relationship follows from disjointness of the emulation (and, hence, of the empirical) dictionaries.
For our {\em coordinated} DDoS with {\em common} emulation dictionary, Theorem~$1$ gives:
\beq
\rho_u=\frac{\alpha}{\alpha+1}, \quad
\rho_{\mathcal{B}_{\textnormal{tot}}}=\frac{\alpha\,B_{\textnormal{tot}}}{\alpha+B_{\textnormal{tot}}}.
\label{eq:ourcase}
\eeq
Notably, the rightmost formulas in~(\ref{eq:singledisjoint}) and~(\ref{eq:ourcase}) reveal that the MIR for the case of disjoint dictionaries  {\em is the same as the MIR of a botnet using a common emulation dictionary}. 
On the other hand, the leftmost formulas in~(\ref{eq:singledisjoint}) and~(\ref{eq:ourcase}) reveal that the MIR of a single bot for the case of disjoint dictionaries {\em is approximately $B_{\textnormal{tot}}$ times smaller than the MIR of a single bot for the case of a common emulation dictionary}. Such a reduced degree of innovation matches the observations reported below~(\ref{eq:EDRdef0}), concerning the flaws of deterministic DDoS attacks based on disjoint emulation dictionaries.  

\noindent
\textsc{Remark VI.}
Assume that the traffic analyst must estimate $\alpha$ based on the patterns collected from a certain subnet $\mathcal{S}$.
From~(\ref{eq:Rfundef}) and~(\ref{eq:theor2}), we have 
%$\alpha=\frac{\lambda_{\mathcal{S}}\,\rho_{\mathcal{S}}}{\lambda_{\mathcal{S}}-\rho_{\mathcal{S}}}$.
$\alpha=
\lambda_{\mathcal{S}}\,\rho_{\mathcal{S}}
/
(\lambda_{\mathcal{S}}-\rho_{\mathcal{S}})
$. 
Accordingly, a reasonable estimator of $\alpha$ can be obtained by replacing $\rho$ and $\lambda$ with their empirical counterparts, yielding: 
\beq
\boxed{
\hat\alpha_{\mathcal{S}}(t)\dfz\frac{\hat\lambda_{\mathcal{S}}(t)\,\hat\rho_{\mathcal{S}}(t)}{\hat\lambda_{\mathcal{S}}(t)-\hat\rho_{\mathcal{S}}(t)}
}
\label{eq:alphaest}
\eeq
In view of Theorem~$1$, such estimator converges in probability to $\alpha$ as $t$ goes to infinity, for any $\mathcal{S}\subseteq\mathcal{B}_{\textnormal{tot}}$.

In contrast, when dealing with normal users, such an interpretation fails in general, since: $i)$ a limiting value $\alpha$ does not necessarily exist, and $ii)$ the generative mechanism of normal patterns is not necessarily interpreted in terms of random picking from an emulation dictionary.
%interpretation of an underlying emulation dictionary used to pick the messages does not necessarily possesses a physical interpretation. 
Nevertheless, the quantity $\hat\alpha_{\mathcal{S}}(t)$ can be meaningfully defined also for arbitrary subnets (i.e., composed also, or even exclusively, by normal users), since it represents the ratio between the empirical rate of ``distinct'' messages $\hat\rho_{\mathcal{S}}(t)$, and the empirical rate of ``repeated'' messages $\hat\lambda_{\mathcal{S}}(t)-\hat\rho_{\mathcal{S}}(t)$, scaled\footnote{The scaling simply corresponds to expressing the result on a per-time-unit basis, rather than on a per-transmission basis.} by the empirical transmission rate $\hat\lambda_{\mathcal{S}}(t)$. 
Such an interpretation is useful since it is now independent from the particular model adopted (transmission scheduling, botnet or normal behavior, etc.).
In the following, even when dealing with arbitrary subnets, we shall loosely refer to $\hat\alpha_{\mathcal{S}}(t)$ as the {\em empirical}, or {\em estimated} EDR.

Finally, exploiting~(\ref{eq:Rfundef}) and~(\ref{eq:alphaest}), the empirical MIR $\hat\rho_{\mathcal{S}}(t)$, for an {\em arbitrary} subnet $\mathcal{S}$, can be expressed as:
\beq
\boxed{
\hat\rho_{\mathcal{S}}(t)=\mathscr{R}(\hat\alpha_{\mathcal{S}}(t),\hat\lambda_{\mathcal{S}}(t))
}
\label{eq:magicformula}
\eeq
%We remark that also the latter expression {\em holds for an arbitrary subnet $\mathcal{S}$}. 

\section{Botnet Identification Condition}
The coordination implied in the {\em distributed} DoS attack introduces some correlation between the empirical dictionaries of the bots, due to the common emulation dictionary where messages are selected.
In contrast, the empirical dictionaries of two normal users are expected to be weakly correlated, due to independence among their activities.
Likewise, the empirical dictionaries of a bot and of a normal user are expected to be weakly correlated, since the network employed by the botmaster to acquire the emulation dictionary is usually not part of the network monitored by the traffic analyst. 

On the other hand, even in the presence of normal (thus, {\em independent}) users, it is realistic to assume a certain degree of {\em physiological} correlation among the users' activities. 
Distinct users can reasonably share parts of their dictionaries, e.g., their surfing activities might partly overlap, due to common interests, popular web-pages, peculiar structure of the destination of interest, etc. 
Similar considerations apply when dealing with a subset of the botnet and a subnet made only of normal users.

Given the very limited amount of information and assumptions we made, and according to the above discussion, any meaningful strategy to discriminate a normal from a malicious behavior, cannot but be based on the degree of dependence among the users. 
In our setting, a convenient way to measure the degree of dependence is provided by the empirical message innovation rate in~(\ref{eq:rhoest}). However, the mere availability of a good network indicator does not provide a {\em quantitative} way to discriminate normal users from bots. In order to design an algorithm for botnet identification, we need to define a proper {\em identification threshold}. To this aim, we can use as reference case for a malicious behavior, the MIR corresponding to the activity performed by a botnet. In order to understand how such operation can be implemented, let us start by considering the case that we must decide whether users $1$ and $2$ belong to a botnet.
Assume for now that the empirical EDRs of the two users obtained through~(\ref{eq:alphaest}) are comparable (the explicit dependence on $t$ being suppressed, for ease of notation, here and in the forthcoming discussion):
\beq
\hat\alpha_1\approx \hat\alpha_2\approx \hat\alpha.
\label{eq:alphassume}
\eeq
When both users belong to a botnet, in view of Theorem~$1$, for $t$ large enough we can write: 
\beqa
\hat\rho_{\{1,2\}}\approx\mathscr{R}(\hat\alpha,\hat\lambda_1+\hat\lambda_2)\dfz\hat\rho_{\textnormal{bot}}.
\label{eq:botapprox}
\eeqa
Moreover, {\em irrespectively of the users' nature}, the empirical MIR of the aggregate subnet $\{1,2\}$ can be upper bounded by the MIR corresponding to disjoint dictionaries, namely,
\beqa
\hat\rho_{\{1,2\}}&\leq&\hat\rho_1+\hat\rho_2
=\mathscr{R}(\hat\alpha_1,\hat\lambda_1)+\mathscr{R}(\hat\alpha_2,\hat\lambda_2)\dfz
\hat\rho_{\textnormal{sum}}\nonumber\\
&\approx&\mathscr{R}(\hat\alpha,\hat\lambda_1)+\mathscr{R}(\hat\alpha,\hat\lambda_2),
\label{eq:rhodisjdef}
\eeqa
where the second equality follows from~(\ref{eq:magicformula}), while the approximate equality follows from~(\ref{eq:alphassume}).
Since from~(\ref{eq:magicbound}) we know that $\hat\rho_{\textnormal{bot}}<\hat\rho_{\textnormal{sum}}$, it makes sense to introduce a threshold lying between the two points $\hat\rho_{\textnormal{bot}}$ and $\hat\rho_{\textnormal{sum}}$, formally, for $\epsilon\in (0,1)$:
\beq
\hat\rho_{\textnormal{bot}}<
\gamma=\hat\rho_{\textnormal{bot}} + \epsilon(\hat\rho_{\textnormal{sum}}-\hat\rho_{\textnormal{bot}})
<
\hat\rho_{\textnormal{sum}}.
\label{eq:gammafirst}
\eeq
When the two users belong to a botnet, from~(\ref{eq:botapprox}) we see that, for large $t$, the empirical MIR $\hat\rho_{\{1,2\}}$ converges to the value $\hat\rho_{\textnormal{bot}}$. 
On the other hand, using Theorem~$1$, it is easy to verify that $\hat\rho_{\textnormal{sum}}-\hat\rho_{\textnormal{bot}}$ converges in probability to a positive quantity, which implies that, for {\em any} $\epsilon>0$, as time elapses, the empirical MIR will stay sooner (higher $\epsilon$) or later (lower $\epsilon$) {\em below} the threshold $\gamma$, yielding:
\beq
\boxed{
\textnormal{$1$ AND $2$ are bots}\Rightarrow \hat\rho_{\{1,2\}}<\gamma
}
\eeq
Consider now the case that at least one user is normal. Were the dictionaries of the two users perfectly disjoint, we would clearly observe, for any $\epsilon\in(0,1)$, that $\hat\rho_{\{1,2\}}\approx \hat\rho_{\textnormal{sum}}>\gamma$. 
However, we already noticed that some correlation is expected to exist even among normal users, or among normal users and bots.
It is also natural to assume that such a correlation is weaker than the correlation exhibited by groups of bots, since the latter are  choosing their messages from one and the same underlying dictionary.\footnote{
In making such assumption, we imply that the specific mechanism used to build normal patterns has a minor influence.} 
%Accordingly, we elect as term of comparison a botnet made of two users, with random message picking, and Poisson time scheduling. The validity of our assumption will be ascertained in the following section when the algorithms will be tested on real network data.   
Accordingly, we might expect that, when at least one user is normal, for sufficiently small $\epsilon$, the empirical MIR still stays above the threshold, namely:
\beq
\boxed{
\textnormal{$1$ OR $2$ are normal}\Rightarrow \hat\rho_{\{1,2\}}>\gamma
}
\label{eq:firstBIC}
\eeq 
In summary, if the empirical MIR stays below $\gamma$, we can declare that the two users form a botnet, otherwise, we can declare that at least one user is normal.

Two main points emerge. 
First, the essential feature enabling a successful discrimination is the assumption in~(\ref{eq:firstBIC}), which accordingly plays the role of a  Botnet Identification Condition (BIC).
Second, the determination of the threshold $\gamma$ relies on a tuning parameter $\epsilon$, which is in principle related to the intrinsic (and unknown) properties of the normal traffic patterns.
Remarkably, the experimental study conducted in the forthcoming Sec.~\ref{sec:expset} will show clearly that: $i)$ the BIC can be safely used, and $ii)$ the choice of $\epsilon$ is by no means critical, even in the non-parametric scenario where no prior information about the normal users' behavior is available. 

Unfortunately, all that glitters is not gold. 
There is an important complication that has been deliberately overlooked so far. 
According to the above explanation, we need to compare the empirical MIR to the MIR of a {\em reference} botnet. 
However, a botnet is characterized by a {\em common} underlying EDR $\alpha$, while in practice we shall typically have $\hat\alpha_1\neq\hat\alpha_2$ (especially when at least one user is normal), implying that the approximation in~(\ref{eq:alphassume}) is unsupported. 
One approach could be that of discarding {\em ab initio} the botnet hypothesis whenever $\hat\alpha_1$ and $\hat\alpha_2$ are too dissimilar. The qualification of being ``too dissimilar'' translates into the appearance of some extra tuning parameter, possibly depending on time, which we want definitely to avoid.
%Since we need to compare the empirical MIR to the MIR of an equivalent botnet, and since a botnet is characterized by a {\em common} underlying EDR $\alpha$, the problem arises of determining a common value from the (distinct) values $\hat\alpha_1$ and $\hat\alpha_2$ observed empirically.
%Clearly, one could define a similarity threshold in such a way that, when $\hat\alpha_1$ and $\hat\alpha_2$ are comparable, the aforementioned procedure is applied. In contrast, when $\hat\alpha_1$ and $\hat\alpha_2$ are enough dissimilar, the hypothesis that users $1$ and $2$ form a botnet could be discarded {\em ab initio}. 
%However, even when faced with a botnet, the empirical values $\hat\alpha_1$ and $\hat\alpha_2$ might differ from the actual EDR $\alpha$, and it might be difficult to define when the dissimilarity is compatible with the botnet hypothesis. The speed of convergence of $\hat\alpha_1$ and $\hat\alpha_2$ toward the true EDR $\alpha$ might be indeed affected by several (unknown) factors. 
%For these reasons, it is advisable to pursue a different strategy.

Another possibility is clearly that of choosing as reference EDR some intermediate value comprised between $\hat\alpha_1$ and $\hat\alpha_2$. In this connection, we remark that the na\"ive choice of the arithmetic average does not work for the following reason. It can be simply verified that, in general, there exist values of $\lambda_1,\lambda_2,\alpha_1,\alpha_2\in\mathbb{R}^+$ for which $\mathscr{R}(\lambda_1,\alpha_1)+\mathscr{R}(\lambda_2,\alpha_2)<\mathscr{R}(\lambda_1+\lambda_2,1/2(\alpha_1+\alpha_2))$, implying that the empirical MIR, {\em even for the case of disjoint dictionaries}, is not necessarily greater than the MIR  of a botnet with reference EDR given by the arithmetic average of $\hat\alpha_1$ and $\hat\alpha_2$. 
A systematic way to select a proper intermediate value is substantially more involved, and is the object of the forthcoming section.

\subsection{Reference EDR by Replacement and Reassignment}
Let us consider two (disjoint) subnets $\mathcal{S}_1$ and $\mathcal{S}_2$, with focus on the case that at least one of them is composed only by normal users, with $\hat\alpha_{\mathcal{S}_1}\neq \hat\alpha_{\mathcal{S}_2}$. 
Recall that we are considering a fixed time $t$, and that the explicit dependence of all quantities upon $t$ is suppressed for ease of notation. 

Since a botnet has {\em common} underlying EDR, and since we want to compare the behavior of $\mathcal{S}_1\cup\mathcal{S}_2$ to that of a botnet, it would be useful to envisage a new pair of traffic patterns for $\mathcal{S}_1$ and $\mathcal{S}_2$ possessing the following characteristics:
\\
\noindent
$i)$ The individual EDRs of $\mathcal{S}_1$ and $\mathcal{S}_2$ are equal, namely (superscript $^\prime$ refers to the ``new'' patterns),
\beq
\hat\alpha^\prime_{\mathcal{S}_1}=\hat\alpha^\prime_{\mathcal{S}_2}=\hat\alpha^\prime.
\label{eq:reassign}
\eeq 
$ii)$ The transmission rate and the MIR of the network $\mathcal{S}_1\cup\mathcal{S}_2$ coincide with those of the original traffic patterns.

We now illustrate a Replacement and Reassignment (RR) procedure, which finds such a new pair starting from the original pattern configuration. Such a procedure relies on the intuitive consideration that, if some messages are reassigned from the subnet with highest EDR to the other subnet, the resulting EDRs tend to keep closer each other. 
In order to avoid misunderstandings, we remark that {\em the RR procedure does not correspond to any real/physical operations made on the traffic patterns. The RR procedure is a conceptual experiment used to demonstrate that it is possible to construct two patterns possessing the aforementioned requirements $i)$ and $ii)$}.

\begin{figure}[t]
\centerline{\includegraphics[width=.6\textheight]{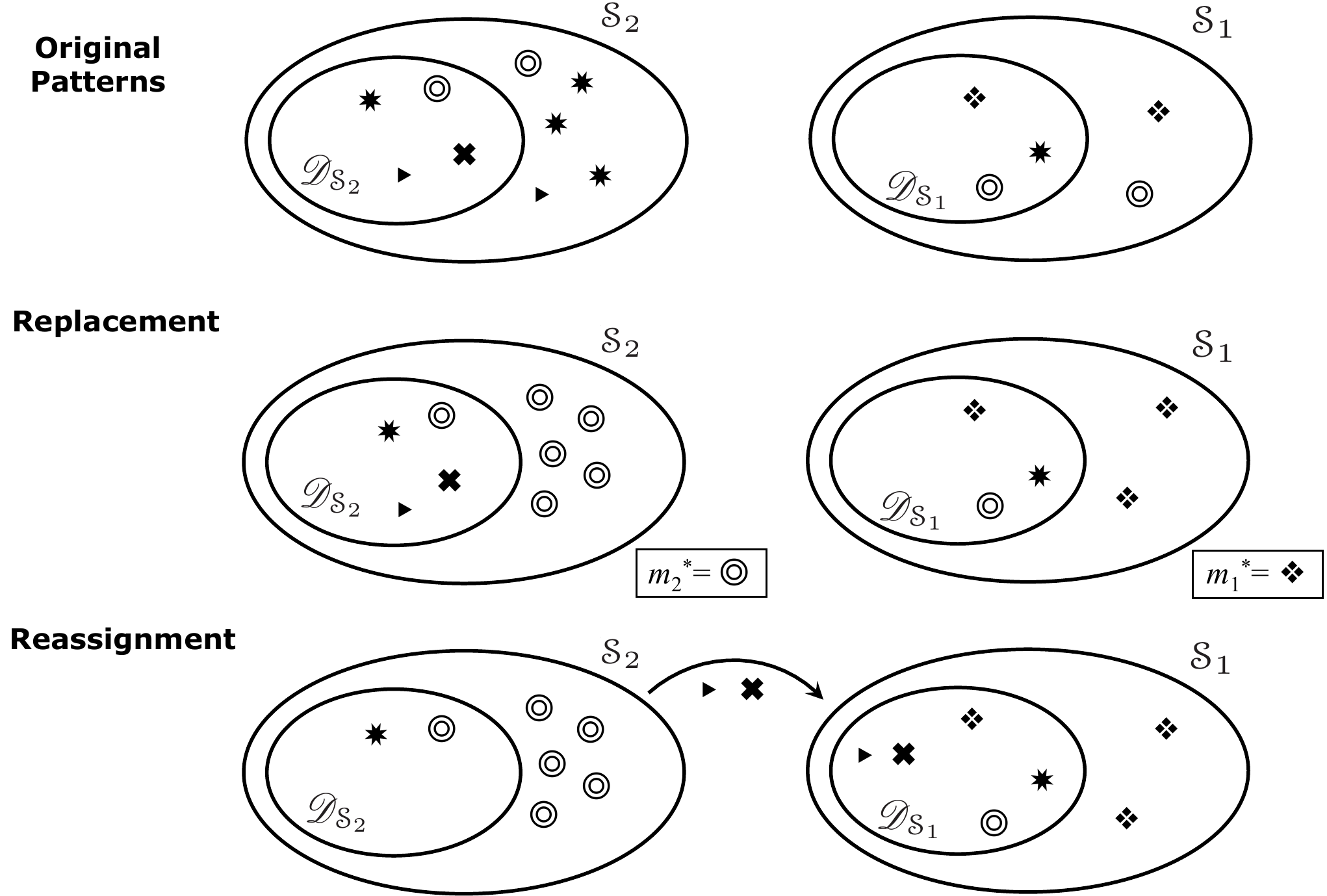}}
\caption{The RR procedure, pictorial exemplification.}
\label{fig:figRR}
\end{figure}

The RR procedure goes as follows --- see Fig.~\ref{fig:figRR} for a pictorial illustration.

\vspace*{1pt}
\noindent
{\em 1. Replacement of repeated messages}. 
The traffic pattern of a subnet $\mathcal{S}$ contains $|\mathscr{D}_{\mathcal{S}}|$ distinct messages, the remaining $N_{\mathcal{S}}-|\mathscr{D}_{\mathcal{S}}|$ ones being repetitions of messages contained in $\mathscr{D}_{\mathcal{S}}$.
The first step of the procedure amounts to replacing such $N_{\mathcal{S}}-|\mathscr{D}_{\mathcal{S}}|$ messages by one and the same message, say it $m^\ast$, contained in $\mathscr{D}_{\mathcal{S}}$. 
The replacement is applied to both subnets $\mathcal{S}_1$ and $\mathcal{S}_2$, with the corresponding replacing messages being $m_1^\ast$ and $m_2^\ast$.  
Since replacement acts only on the message content, the transmission rates do not change. 
Moreover, since replacement leaves unaltered the number of distinct messages within each subnet, the MIR of the subnets, and the MIR of $\mathcal{S}_1\cup\mathcal{S}_2$, are unaltered.\footnote{The MIR is determined only by the content of the empirical dictionaries.}

\vspace*{1pt}
\noindent
{\em 2. Reassignment of messages}.
%%%
Some messages will be reassigned from one subnet to the other subnet (only in one direction, namely, either from $\mathcal{S}_2$ to $\mathcal{S}_1$ or from $\mathcal{S}_1$ to $\mathcal{S}_2$). 
For the sake of clarity, assume that $\mathcal{S}_2$ is ``passing'' some of its messages to $\mathcal{S}_1$, with the prescription that the replacing message $m^\ast_2$ is never passed. 
Since, after replacement, all messages different from $m^\ast_2$ appear only once in the pattern of $\mathcal{S}_2$, we see that {\em all} messages passed to $\mathcal{S}_1$ are necessarily distinct.
The rate of messages (number of messages normalized to the current time $t$) that are reassigned from $\mathcal{S}_2$ to $\mathcal{S}_1$ is denoted by $\Delta$. Accordingly, a negative $\Delta$ will correspond to the converse situation where $\mathcal{S}_1$ passes some of its messages to $\mathcal{S}_2$. 
As a result, the transmission rates of the pattern configuration after reassignment are:
\beq
(\hat\lambda^\prime_{\mathcal{S}_1},\hat\lambda^\prime_{\mathcal{S}_2})=(\hat\lambda_{\mathcal{S}_1}+\Delta,\hat\lambda_{\mathcal{S}_2}-\Delta).
\label{eq:lamnewpattern}
\eeq
Moreover, since the correlation between the two patterns is weak (recall that one of the subnets is composed only by normal users), we assume that it is always possible to reassign messages that do not belong to the intersection of the two empirical dictionaries. 
Such assumption, along with the fact that all passed messages are distinct, implies that, in terms of individual MIRs, what is lost by a subnet is exactly gained by the other subnet. Formally:
\beq
(\hat\rho^\prime_{\mathcal{S}_1},\hat\rho^\prime_{\mathcal{S}_2})=(\hat\rho_{\mathcal{S}_1}+\Delta,\hat\rho_{\mathcal{S}_2}-\Delta).
\label{eq:rhonewpattern}
\eeq
Note that not all values of $\Delta$ are admissible. For instance, if messages from $\mathcal{S}_2$ are reassigned to $\mathcal{S}_1$, the rate of reassigned messages cannot exceed the rate of distinct messages owned by $\mathcal{S}_2$, namely, $\Delta\leq\hat\rho_{\mathcal{S}_2}$. Likewise, in the converse case, $-\Delta\leq\hat\rho_{\mathcal{S}_1}$, finally yielding:\footnote{
Actually, since we exclude the replacing messages $m^\ast_1$ or $m^\ast_2$ from the reassignment procedure, a subnet cannot pass all its distinct messages. However, for large $t$ the contribution of a {\em single} message becomes irrelevant. }
\beq
-\hat\rho_{\mathcal{S}_1}
\leq
\Delta
\leq
\hat\rho_{\mathcal{S}_2}.
\label{eq:admisscond}
\eeq
Moreover, since the reassignment changes only the ``owner'' of a given message, the MIR of the {\em aggregate} network $\mathcal{S}_1\cup\mathcal{S}_2$ is left unaltered, namely, $\hat\rho^\prime_{\mathcal{S}_1\cup\mathcal{S}_2}=\hat\rho_{\mathcal{S}_1\cup\mathcal{S}_2}$.

\vspace*{1pt}
\noindent
{\em 3. Choice of $\Delta$ for the equilibrium condition}.
At the end of the reassignment procedure, the new EDRs corresponding to $\mathcal{S}_1$ and $\mathcal{S}_2$ become, respectively,
$
\hat\alpha^\prime_{\mathcal{S}_1}=\hat\lambda_{\mathcal{S}_1}^\prime \hat\rho_{\mathcal{S}_1}^\prime
(
\hat\lambda_{\mathcal{S}_1}^\prime-\hat\rho_{\mathcal{S}_1}^\prime
)$, and 
$
\hat\alpha^\prime_{\mathcal{S}_2}=\hat\lambda_{\mathcal{S}_2}^\prime \hat\rho_{\mathcal{S}_2}^\prime
/
(\hat\lambda_{\mathcal{S}_2}^\prime-\hat\rho_{\mathcal{S}_2}^\prime) 
$, where we have exploited~(\ref{eq:alphaest}). 
In order to get a common reference EDR $\hat\alpha^\prime$, we enforce the condition in~(\ref{eq:reassign}), which, using~(\ref{eq:lamnewpattern}) and~(\ref{eq:rhonewpattern}) into the latter two equations, amounts to seek a value $\Delta^\star$ such that:
\beq
\hat\alpha^\prime=
\frac{(\hat\lambda_{\mathcal{S}_1}+\Delta^\star)(\hat\rho_{\mathcal{S}_1}+\Delta^\star)}{\hat\lambda_{\mathcal{S}_1}-\hat\rho_{\mathcal{S}_1}}
=
\frac{(\hat\lambda_{\mathcal{S}_2}-\Delta^\star)(\hat\rho_{\mathcal{S}_2}-\Delta^\star)}{\hat\lambda_{\mathcal{S}_2}-\hat\rho_{\mathcal{S}_2}},
\label{eq:alphaquad}
\eeq
with the additional prescription that condition~(\ref{eq:admisscond}) is met.
%In Appendix~\ref{app:B} it is shown that the above quadratic equation in fact admits such a solution.
Therefore, the explicit formula for $\Delta^\star$ is found by solving a quadratic equation, and by simple algebra it can be verified that the solution fulfilling~(\ref{eq:admisscond}) is:
\beqa
\Delta^\star&=&\frac{
\hat\lambda_{\mathcal{S}_1}\hat\lambda_{\mathcal{S}_2}
-
\hat\rho_{\mathcal{S}_1}\hat\rho_{\mathcal{S}_2}
}
{
{(\hat\lambda_{\mathcal{S}_1}-\hat\rho_{\mathcal{S}_1})
-
(\hat\lambda_{\mathcal{S}_2}-\hat\rho_{\mathcal{S}_2})}
}
\nonumber\\
&-&
\frac{\sqrt{(\hat\lambda_{\mathcal{S}_1}-\hat\rho_{\mathcal{S}_1})(\hat\lambda_{\mathcal{S}_2}-\hat\rho_{\mathcal{S}_2})(\hat\lambda_{\mathcal{S}_1}+\hat\rho_{\mathcal{S}_2})(\hat\lambda_{\mathcal{S}_2}+\hat\rho_{\mathcal{S}_1})}
}
{
{(\hat\lambda_{\mathcal{S}_1}-\hat\rho_{\mathcal{S}_1})
-
(\hat\lambda_{\mathcal{S}_2}-\hat\rho_{\mathcal{S}_2})}
}
.\nonumber\\
\eeqa
From~(\ref{eq:alphaquad}), it is easily verified that a positive $\Delta^\star$ corresponds to $\hat\alpha_{\mathcal{S}_1}<\hat\alpha^\prime<\hat\alpha_{\mathcal{S}_2}$ (while the latter two inequalities are reversed when $\Delta^\star<0$), implying that the subnet with the highest EDR ``passes'' a fraction of its messages to the other subnet. In summary, we conclude that:
\beq
\boxed{
\min(\hat\alpha_{\mathcal{S}_1},\hat\alpha_{\mathcal{S}_2})
\leq
\hat\alpha^\prime
\leq
\max(\hat\alpha_{\mathcal{S}_1},\hat\alpha_{\mathcal{S}_2})
}
\label{eq:alphamaxmin}
\eeq
%Since the overall activity of $\mathcal{S}_1\cup\mathcal{S}_2$ is left unchanged by replacement and reassignment, it makes sense to compare such a MIR to the MIR corresponding to the case that $\mathcal{S}_1$ and $\mathcal{S}_2$ form a botnet, where the reference value of the EDR is now consistently chosen as $\hat\alpha^\prime$. 
According to the above explanation, when at least one of the subnets is composed only by normal users, we can write: 
\beqa
\hat\rho_{\textnormal{sum}}(\mathcal{S}_1,\mathcal{S}_2)
&\dfz&
\hat\rho_{\mathcal{S}_1}+\hat\rho_{\mathcal{S}_2}
\stackrel{\textnormal{(a)}}{=}
\hat\rho^\prime_{\mathcal{S}_1}+\hat\rho^\prime_{\mathcal{S}_2}
\nonumber\\
&\stackrel{\textnormal{(b)}}{=}&
\mathscr{R}(\hat\alpha^\prime,\hat\lambda^\prime_{\mathcal{S}_1})
+\mathscr{R}(\hat\alpha^\prime,\hat\lambda^\prime_{\mathcal{S}_2})
\nonumber\\
&\stackrel{\textnormal{(c)}}{>}&
\mathscr{R}(\hat\alpha^\prime,\hat\lambda^\prime_{\mathcal{S}_1}+\hat\lambda^\prime_{\mathcal{S}_2})\nonumber\\
&\stackrel{\textnormal{(d)}}{=}&\mathscr{R}(\hat\alpha^\prime,\hat\lambda_{\mathcal{S}_1}
+\hat\lambda_{\mathcal{S}_2})
\dfz
\hat\rho_{\textnormal{bot}}(\mathcal{S}_1,\mathcal{S}_2),
\label{eq:fundamentalmagicbound}
\eeqa 
where $(a)$ follows from~(\ref{eq:rhonewpattern}); $(b)$ follows from~(\ref{eq:magicformula}); $(c)$ follows from~(\ref{eq:magicbound}); and $(d)$ follows from~(\ref{eq:lamnewpattern}).
On the other hand, when $\mathcal{S}_1$ and $\mathcal{S}_2$ form a botnet, Theorem~$1$ implies that, for $t$ large enough, $\hat\alpha_{\mathcal{S}_1}\approx\hat\alpha_{\mathcal{S}_2}\approx\alpha$, which in turn implies that $\hat\alpha^\prime\approx\alpha$ in view of~(\ref{eq:alphamaxmin}). Therefore, in this case the inequality 
$\hat\rho_{\textnormal{sum}}(\mathcal{S}_1,\mathcal{S}_2)>\hat\rho_{\textnormal{bot}}(\mathcal{S}_1,\mathcal{S}_2)$ is justified by the approximations: 
$\hat\rho_{\textnormal{sum}}(\mathcal{S}_1,\mathcal{S}_2)
\approx
\mathscr{R}(\alpha,\lambda_{\mathcal{S}_1})+\mathscr{R}(\alpha,\lambda_{\mathcal{S}_2})
$ and 
$\hat\rho_{\textnormal{bot}}(\mathcal{S}_1,\mathcal{S}_2)
\approx
\mathscr{R}(\alpha,\lambda_{\mathcal{S}_1}+\lambda_{\mathcal{S}_2})
$.

We have in fact shown that, for {\em arbitrary transmission schedulings and message-picking policies}, the empirical MIR of a botnet with reference EDR value~(\ref{eq:alphaquad}) does {\em always} provide a lower bound to the sum of individual MIRs.\footnote{We remark that the aforementioned result does not relate in any way to the deterministic or Poisson scheduling and to the random message picking that characterize the class of DDoS attacks considered in the present work.}

\begin{figure*}[t]
\centerline{\includegraphics[width=.7\textheight]{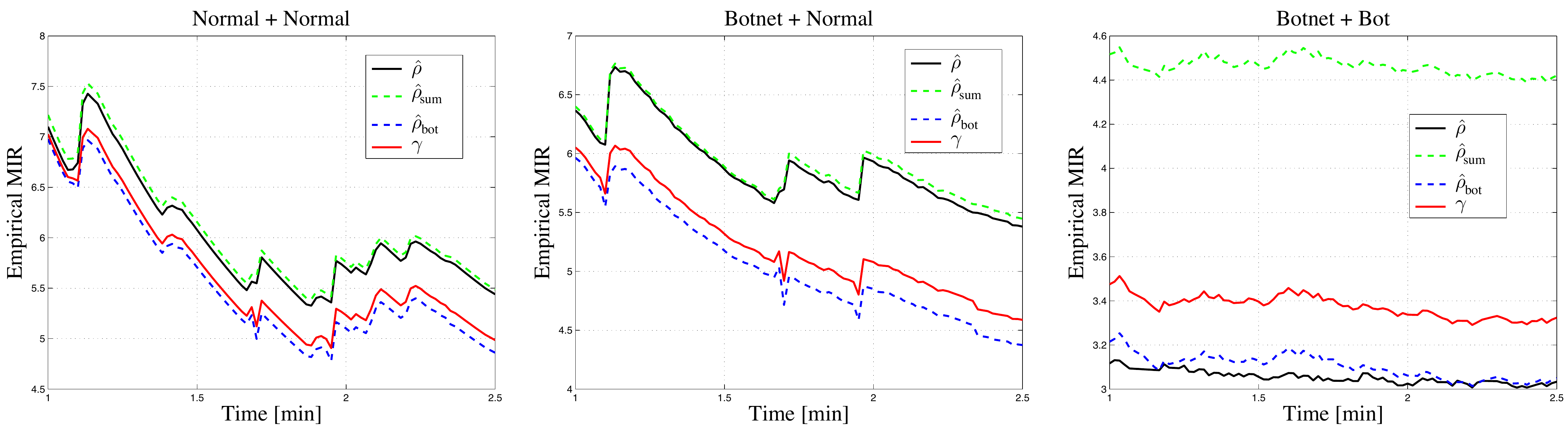}}
\caption{Time evolution of the empirical message innovation rate $\hat\rho$ (solid, black), compared to the identification threshold $\gamma$ (solid, red). For comparison purposes, the upper bound corresponding to the case of disjoint dictionaries, $\hat\rho_{\textnormal{sum}}$ (dashed, green), and the lower bound corresponding to the botnet case, $\hat\rho_{\textnormal{bot}}$ (dashed, magenta) are displayed. Moving from left to right, the different panels refer to $i)$ the union of two normal users; $ii)$ the union of a botnet of size $10$ and a normal user; and $iii)$ the union of a botnet of size $10$ and a bot.}
\label{fig:fig0}
\end{figure*}

\subsection{Threshold Setting}
Let us introduce an intermediate threshold lying between the lower bound and the upper bound in~(\ref{eq:fundamentalmagicbound}), namely, for $\epsilon\in(0,1)$, 
\beq
\boxed{
\gamma(\mathcal{S}_1,\mathcal{S}_2)=\hat\rho_{\textnormal{bot}}(\mathcal{S}_1,\mathcal{S}_2)+
\epsilon\,[\hat\rho_{\textnormal{sum}}(\mathcal{S}_1,\mathcal{S}_2)-\hat\rho_{\textnormal{bot}}(\mathcal{S}_1,\mathcal{S}_2)]
}
\label{eq:gammagamma}
\eeq 
When $\mathcal{S}_1$ and $\mathcal{S}_2$ form a botnet, from Theorem~$1$ it is immediately seen (recall that $\hat\alpha^\prime$ will converge to the true $\alpha$) that $\hat\rho_{\mathcal{S}_1\cup\mathcal{S}_2}<\gamma(\mathcal{S}_1,\mathcal{S}_2)$ as $t\rightarrow\infty$.

When at least one of the subnets is made of normal users, the degree of dependence among their patterns is low. 
Since $i)$ we have shown that there exist two patterns, {\em with common EDR}, $\hat\alpha^\prime$, and with the same {\em joint} properties (overall transmission rate and MIR) of the original patterns; and $ii)$ the RR procedure only replaces and/or reassigns messages, it is expected that the joint MIR of a botnet with EDR $\hat\alpha^\prime$ is lower than $\hat\rho_{\mathcal{S}_1\cup\mathcal{S}_2}$.
Otherwise stated, it is reasonable to assume that $\hat\rho_{\mathcal{S}_1\cup\mathcal{S}_2}$, even if not coinciding with the upper bound $\hat\rho_{\textnormal{sum}}(\mathcal{S}_1,\mathcal{S}_2)$ in~(\ref{eq:fundamentalmagicbound}), is still sufficiently far from the lower bound $\hat\rho_{\textnormal{bot}}(\mathcal{S}_1,\mathcal{S}_2)$. These considerations, for small $\epsilon$, implicitly define the following identification condition.

\noindent
{\bf Botnet Identification Condition (BIC)}

\noindent
Let $\mathcal{S}_1$ and $\mathcal{S}_2$ be two subnets with $\mathcal{S}_1\bigcap\mathcal{S}_2=\emptyset$. 
{\em If at least one of the subnets is composed only by normal users}:
\beq
\boxed{
\hat\rho_{\mathcal{S}_1\cup\mathcal{S}_2}\geq\gamma(\mathcal{S}_1,\mathcal{S}_2)
}
\label{eq:magiclowerboundfinal}
\eeq
We remark that the case of $\mathcal{S}_1$ arbitrary vs. $\mathcal{S}_2$ arbitrary is not dealt with. This is not unintentional, since, as it will be clear from Theorem~$2$, the two situations discussed are sufficient to devise a {\em consistent} botnet identification algorithm.

In summary, we end up with the following recipe:
\beqa
\hspace*{-20pt}
\textnormal{{\footnotesize $\mathcal{S}_1$ AND $\mathcal{S}_2$ contain only bots}}&\Rightarrow& \hat\rho_{\mathcal{S}_1\cup\mathcal{S}_2}<\gamma(\mathcal{S}_1,\mathcal{S}_1),
\label{eq:below}\\
\hspace*{-20pt}
\textnormal{{\footnotesize $\mathcal{S}_1$ OR $\mathcal{S}_2$ contain only normal users}}&\Rightarrow& \hat\rho_{\mathcal{S}_1\cup\mathcal{S}_2}\geq\gamma(\mathcal{S}_1,\mathcal{S}_1).
\label{eq:above}
\eeqa
In Fig.~\ref{fig:fig0}, we illustrate the significance of the BIC. The normal users' activity refers to a monitoring campaign conducted over real data. The bots' activity has been generated according to the model described in Sec.~\ref{sec:bibiebibotnet}.
The details of such a campaign will be given in the forthcoming section. In all the three panels we display: the empirical MIR, the threshold $\gamma$ in~(\ref{eq:gammagamma}), along with its upper ($\hat\rho_{\textnormal{sum}}$) and lower ($\hat\rho_{\textnormal{bot}}$)  bounds.
An observation window of $2.5$ min is considered. All the relevant quantities are updated each $1$ s, and both quantities are displayed as functions of time, in the interval between $1$ and $2.5$ min.

In the leftmost panel, we address the case of a pair of normal users. 
We see that the MIR stays (slightly) below the upper bound, meaning that a certain degree of correlation exists. However, {\em the MIR stands clear above the threshold, as prescribed by~(\ref{eq:above}), and confirming the validity of the BIC}.

In the middle panel, the two subnets under test, $\mathcal{S}_1$ and $\mathcal{S}_2$, are a botnet of size $10$, and a normal user, respectively. 
Conclusions similar to those pertaining to a normal-normal pairing can be drawn, substantiating again the BIC.
We further see that, at the beginning of the observation window, the activities of the two subnets are almost independent, i.e., the MIR essentially matches the upper bound. As time elapses, a certain degree of correlation appears, {\em but the MIR still stays above the threshold}.
 
Finally, in the rightmost panel, the case of a botnet/bot interaction is addressed. We see that the empirical MIR: $i)$ approaches, as time elapses, the quantity $\hat\rho_{\textnormal{bot}}$, in perfect agreement with Theorem~1, and $ii)$ stands clear below the threshold, in perfect agreement with~(\ref{eq:below}).

In summary, the picture obtained from the above analysis reveals that the theoretical findings of Theorem~1, as well as the conjectured behavior of the normal users implied by the BIC, are confirmed {\em over real network traces}.\footnote{Needless to say, our experiments have been repeated for many pairs of  normal users, not reported here for obvious reasons.}
%%%%%%%%%%%%%%%%%%%%%%%%%%%%%%%%

\section{The BotBuster Algorithm}
\label{sec:algo}
We now focus on the derivation of the inference algorithm aimed at disclosing a botnet possibly hidden in the network. 
The {\em BotBuster} algorithm is described by the pseudo-code reported in the right column above, and basically exploits the fact that, given two disjoint subnets, the BIC allows to discriminate the situation where both subnets are part of a botnet, from the situation where at least one of them is made of normal users. 
We shall show that the proposed algorithm possesses the fundamental requirement of {\em consistency}, namely, {\em the guarantee that the botnet is correctly identified as $t$ grows}.

Let us examine how the algorithm works. 
First, note that a botnet made of one user, besides making little sense in practice, is by definition non-identifiable, since we assumed that the characteristics of the messages at a single-user level do not reveal any special information.
Now, at the beginning of the algorithm, user $1$ is initially declared as a bot, namely, $\hat{\mathcal{B}}=\{1\}$.
Then, it is checked whether users $1$ and $2$ form a botnet. If so, $\hat{\mathcal{B}}=\{1,2\}$ is taken as the current botnet estimate. If not, $\hat{\mathcal{B}}=\{1\}$ is retained. 
Then, it is checked whether the currently estimated botnet $\hat{\mathcal{B}}$ forms a bot with user $3$, and so on.
At the end of the inner loop, the algorithm ends up with an estimate $\hat{\mathcal{B}}$. If the cardinality of the estimated set is greater than one, it is taken as a current estimate. 

The procedure is then restarted by choosing user $2$ as initial pivot, and sequentially checking the remaining users as explained before. 
At the end of the inner loop, the algorithm ends up with another estimate $\hat{\mathcal{B}}$. If the cardinality of the estimated set is greater than one {\em and} greater than the cardinality of the previously estimated set\footnote{When $t$ is large and the BIC is {\em perfectly} verified, the inner loop ends with either an empty set or the true botnet. Thus, selecting the estimate with the highest cardinality might appear redundant. 
Such operation is instead useful when operating under non-ideal conditions, as we shall explain soon.}, then it is taken as a current estimate. Otherwise, the previous estimate is retained. 
The procedure ends when all users have been scanned as pivots.

%%%
\RestyleAlgo{boxruled}
\begin{algorithm}[t]
\caption{
\vspace*{5pt}
$\hat{\mathcal B}_{\textnormal{new}}$=BotBuster
}
$\mathcal{N}=\{1,2,\dots,N\}$; $\hat{\mathcal B}_{\textnormal{new}}=\emptyset$\;
\For{$b_0\in\mathcal{N}$}{
$\hat{\mathcal B}=\{b_0\}$\;
\For{$j\in\mathcal{N}\setminus\{b_0\}$}{
	\If{$
	\hat\rho(\hat{\mathcal{B}}\cup\{j\})<\gamma(\hat{\mathcal{B}},\{j\})
	$}{
	$\hat{\mathcal{B}}=\hat{\mathcal{B}}\bigcup\{j\}$\;
	}
}
	\If{$|\hat{\mathcal B}|>\max(1,|\hat{\mathcal{B}}_{\textnormal{new}}|)$}{
	$\hat{\mathcal{B}}_{\textnormal{new}}=\hat{\mathcal{B}}$\;
	}
}
\end{algorithm}

We see that, under the BIC, all checks performed by the algorithm will give eventually the right answer, with probability tending to $1$ as $t\rightarrow\infty$. BotBuster is accordingly expected to provide a {\em consistent} botnet estimator, as will be stated and proved in the forthcoming Theorem~$2$.
The algorithm complexity is $\mathcal{O}(N^2)$ (only pairwise checks are performed), which is definitely tolerable, since we are seeking, within a network of size $N$, a subset of unknown size that matches some prescribed conditions. 
%The complexity of the single checking step amounts to the evaluation of the estimated MIR, and of the threshold function $\gamma(\mathcal{S}_1,\mathcal{S}_2)$ in~(\ref{eq:gammagamma}). Accordingly, such complexity changes depending on the realization, since it is related to the size of the botnet estimated at a given step. 
Finally, the looping structure of the algorithm makes it naturally open to parallelization, which is especially important for large networks.

%\subsection{Inference Performance of BotBuster}
In order to quantify the algorithm performance, we need to choose some meaningful indicators. 
With reference to a network $\mathcal{N}=\{1,2,\dots,N\}$, containing a botnet $\mathcal{B}$, and letting $\hat{\mathcal B}(t)$ be the botnet estimated at time $t$ by BotBuster, we introduce the following performance indices:
\beq
\eta_{\textnormal{bot}}(t)=\frac{\E[|\hat{\mathcal{B}}(t)\cap\mathcal{B}|]}{|\mathcal{B}|},\quad
\eta_{\textnormal{nor}}(t)=\frac{\E[|\hat{\mathcal{B}}(t)\cap(\mathcal{N}\setminus\mathcal{B})|]}{|\mathcal{N}\setminus\mathcal{B}|},
\eeq
namely, the expected fraction of {\em correctly banned users} (i.e., discovered bots), and the expected fraction of incorrectly-banned users (i.e., normal users erroneously declared as bots). Clearly, $\eta_{\textnormal{bot}}(t)$ (resp., $\eta_{\textnormal{nor}}(t)$) is not defined when $\mathcal{B}=\emptyset$ (resp., when $\mathcal{B}=\mathcal{N}$). 
We would like to see $\eta_{\textnormal{bot}}(t)\rightarrow 1$, and $\eta_{\textnormal{nor}}(t)\rightarrow 0$ as $t$ goes to infinity. 
Under the ideal assumption that the BIC is always verified, such requirement is in fact fulfilled, as stated in the following theorem.

\noindent
{\bf \textsc{Theorem 2} (Consistency of BotBuster).}
{\em 
Consider a network $\mathcal{N}=\{1,2,\dots,N\}$, containing a botnet $\mathcal{B}$, with $|\mathcal{B}|\neq 1$, launching a randomized DDoS attack. 
The bots' transmission policies are either synchronous with constant transmission rate, or independent Poisson processes, while the normal users' transmission policies are arbitrary.
%Let $\hat{\mathcal B}(t)$ be the botnet estimated at time $t$ by the algorithm BotBuster. 
Then, for any finite emulation dictionary rate $\alpha$, the algorithm BotBuster is consistent, namely,
\beq
\boxed{
\lim_{t\rightarrow\infty}\eta_{\textnormal{bot}}(t)=1,\qquad
\lim_{t\rightarrow\infty}\eta_{\textnormal{nor}}(t)=0
}
\label{eq:etabotnorconv}
\eeq
The claim for the case $\mathcal{B}=\emptyset$ (resp., $\mathcal{B}=\mathcal{N}$) is intended to hold with reference solely to $\eta_{\textnormal{nor}}(t)$ (resp., to $\eta_{\textnormal{bot}}(t)$).     
}

{\em Proof}. See Appendix~\ref{app:C}.~$\hfill\blacksquare$

Theorem $2$ reveals that the botnet estimated by BotBuster converges to the {\em true} one as time elapses. 
The fundamental requirement enabling such result is the BIC validity. 
On the other hand, in real-world applications, the assumption that the BIC is verified {\em for all} normal/normal and botnet/normal interactions, as well as {\em for all} time epochs, is surely an {\em over-idealized} one. 
It cannot be excluded that, occasionally, two independent users feature an unusual degree of superposition between their empirical dictionaries, giving rise to spurious clusters of normal users that might be erroneously included in the estimated botnet. 
What is expected to be true even in real-world applications, is that such cases are rare and that the clusters' cardinality is small. 
%Since we need solutions that might find application, it is important to understand how performance changes under these non-ideal conditions. 
Now, since the algorithm selects the estimate $\hat{\mathcal{B}}$ with the {\em highest} cardinality, and since {\em distributed} DoS attacks with small botnet sizes make little sense, estimated botnets of unreasonably small cardinality should be easily ruled out by BotBuster. 
As a result, the final estimate is likely to contain the true botnet, plus (possibly) a small fraction of normal users. 
Thus, {\em even under non-ideal operation conditions}, it is expected that $\eta_{\textnormal{bot}}(t)\rightarrow 1$ as $t\rightarrow\infty$, whereas $\eta_{\textnormal{nor}}(t)$ possibly takes on some small value.

%We finally remark that in our genuine estimation context, the focus is not on {\em detecting} the DDoS attack, and the algorithm is tasked to {\em estimate} a (reasonably sized) botnet assumed to be present in the network. 
%Nevertheless, it is useful to see how the algorithm is expected to work when there is {\em no} botnet hidden in the network.
%According to the above explanation, in this case BotBuster is likely to output an estimate with relatively small cardinality, implying that $\eta_{\textnormal{nor}}(t)$ takes on some relatively small value. 
%Now, depending on the particular policy adopted by the network manager, one could either $i)$ declare that no attack is occurring, since the estimated botnet is not compatible with the anomalous request rate that triggered the identification procedure, or, $ii)$ at worst, pay the price of banning a small fraction of normal users. 

%%%%%%%%%%%%%%%%%%%%%%%%%%%%%%%%
\section{Experimental Results}
\label{sec:expset}
%The analysis conducted so far relies on {\em theoretical} results.
%Therefore, it is necessary to test the applicability of such results to practical problems. 
%To this aim, we now proceed in three basic steps: $i)$ we detail the experimental setting adopted for collecting {\em real} network traces; $ii)$ we illustrate the procedure adopted to build the bots' patterns; $iii)$ we examine the performance BotBuster in the pertinent experimental setting. 
%\subsection{Network Traces Collection} 
The theoretical analysis conducted in the previous sections relies upon a number of assumptions. 
As a result, when dealing with {\em real} network traces, and with challenging DDoS attacks, the operational validity of the algorithm BotBuster is not at all obvious. 
This is why we have performed a detailed experimental analysis, whose outcomes are now reported.

A popular e-commerce website has been selected as target destination of the attack. 
Clearly, the normal users have no attacking intent, they perform ordinary surfing activity. 
About $20$ min of (application-layer) traffic have been collected, from $10$ independent users, which were students and researchers working in our laboratory, and carrying on their surfing activity almost independently.
The collected streams have been partitioned into chunks of $2$ min. 
In the forthcoming analysis we take two perspectives. 
In one scenario, the number of normal users is $10$, and, per each trial, we choose $10$ traces of $2$ min, at random among the available traces.
In the other scenario, 2-min chunks belonging to the {\em same} user have been treated as if they were coming from {\em distinct} users. In this way, we multiplied (fictitiously) the number of available normal users. 
We stress that considering each chunk as an independent user is an approximation, but at the same time such approximation introduces a spurious correlation among the users, which clearly makes more challenging the botnet identification.

%\subsection{Building the Botnet}
The DDoS attack has been generated so as to fall into the class described in Sec.~\ref{sec:bibiebibotnet}. 
Given the dictionary of messages obtained from the {\em whole} activity recorded in the laboratory, it is assumed that, at epoch $t$, only the first $\lfloor e_0+\alpha t \rfloor$ messages of such a dictionary are available to the botnet, giving rise to the emulation dictionary $\mathscr{E}(t)$, for fixed parameters $e_0$ (size of the dictionary at $t=0$) and $\alpha$. 
Then, independently at each bot, a Poisson time-scheduling is randomly generated, and, per each transmission epoch $t$, each bot picks messages at random from the currently available $\mathscr{E}(t)$.
\begin{figure}[t]
\centerline{\includegraphics[width=.45\textheight]{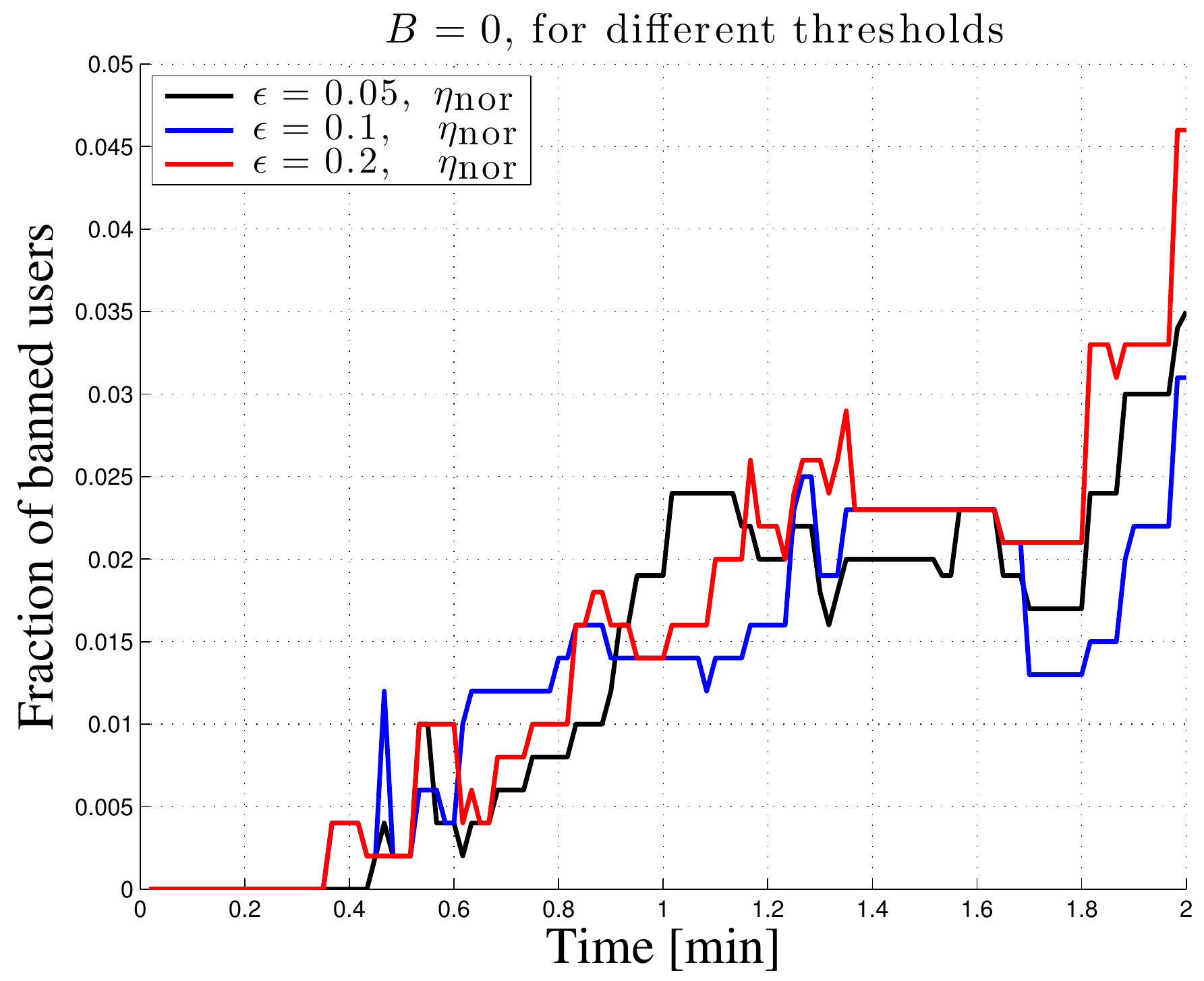}}
\caption{Fraction of banned users as a function of time, for different values of the threshold parameter $\epsilon$. The monitored network is composed of $10$ normal users, and contains no bots. 
The depicted curves are computed over $100$ independent Monte Carlo trials. Per each trial, $2$-min chunks of each user are randomly selected among the available chunks.}
\label{fig:fig1}
\end{figure}

%\subsection{The Algorithm at Work}
We are now ready to examine the performance of our algorithm. 
We recall that our algorithm is non-parametric, namely, that it does assume knowledge neither of the transmission rates, nor of the parameters of the botnet emulation dictionary ($e_0$ and $\alpha$). In contrast, the size of the network is obviously known. 
The only input parameter is the factor $\epsilon$ appearing into~(\ref{eq:gammagamma}). 
Our experiments will serve to ascertain:
\begin{itemize}
\item
The practical significance of the BIC over {\em real} data.
\item
The ability of BotBuster in discovering the hidden bots.
\item
The role of the algorithm parameters.
\item
The role of the botnet power (the learning ability $\alpha$).
\end{itemize}
In Fig.~\ref{fig:fig1}, we consider a network made of $10$ normal users, without bots. 
Therefore, the quantity $\eta_{\textnormal{bot}}$ makes little sense, and is accordingly not displayed. 
The observation window lasts $2$ min, and the simulation points refer to the output of the algorithm taken each $1$ s. 
The algorithm is run for three values of the threshold parameter $\epsilon\in(0,1)$, namely, $0.05, 0.1,$ and $0.2$, and the estimates are averaged over $100$ Monte Carlo trials.
Now, were the BIC exactly verified for any subset of normal users, and for any time epoch, the fraction of banned users should be always zero. As already discussed, in practice the BIC is expected to be verified approximately. 
This notwithstanding, in Fig.~\ref{fig:fig1} we see that the percentage of erroneously banned users is very small for all the thresholds in the considered range, never exceeding $5\%$. Notably, such behavior suggests that a BIC violation is unlikely to occur, and that, in any case, it involves small groups of users.

\begin{figure}[t]
\centerline{\includegraphics[width=.45\textheight]{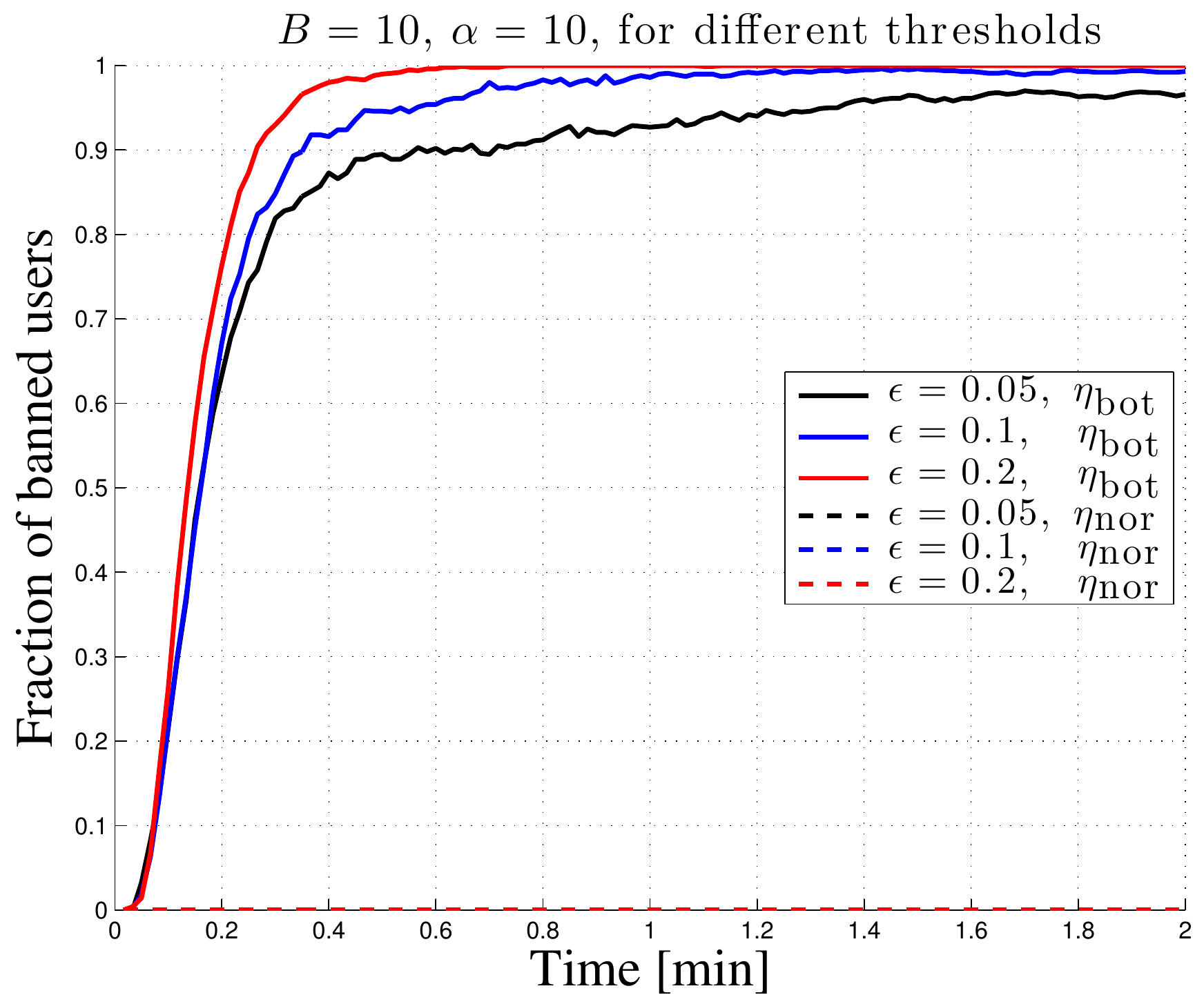}}
\caption{Fraction of banned users as a function of time, for different values of the threshold parameter $\epsilon$. The monitored network is composed of $10$ normal users, and contains $B=10$ bots. 
Solid curves refer to correctly banned bots, whereas dashed curves refer to erroneously banned normal users.
The depicted curves are computed over $100$ independent Monte Carlo trials. Per each trial, $2$-min chunks of each user are randomly selected among the available chunks.}
\label{fig:fig2}
\end{figure}

In Fig.~\ref{fig:fig2}, the same analysis is repeated for the case that the network comprises $10$ normal users plus $10$ bots. The botnet EDR is $\alpha=10$. We remark that such a value is compatible with some of the empirical values $\hat\alpha$ estimated over the normal users' traces.
%, which implies that no particular inference could be made by simply examining the single-user behavior. 
%, showing then a markedly different behavior (the algorithm basically stops working properly) for the higher value $\epsilon=1$. 
We see that the dashed curves are in practice invisible, revealing that the estimated $\eta_{\textnormal{nor}}$ is almost zero for all the considered values of $\epsilon$.
%, offering then a scarce performance for the higher value $\epsilon=1$. 
This behavior should be contrasted to what observed in Fig.~\ref{fig:fig1}, where, in the absence of a botnet, the BIC was occasionally violated. However, as discussed at the end of Sec.~\ref{sec:algo}, the spurious-and-small estimated clusters containing normal users can be efficiently ruled out by the fact that the algorithm selects, as a final estimate, only the cluster with maximum size, which is expected to contain only bots. 

With regard to the fraction of correctly identified bots, we see that $\eta_{\textnormal{bot}}$ increases as $\epsilon$ increases from $0.05$ to $0.2$. In fact, increasing $\epsilon$ makes it easier staying {\em below} the threshold, which facilitates the inclusion of a node in the estimated botnet. 

%\textcolor{blue}{
%On the other hand, the BIC is expected to fail beyond a certain $\epsilon$, since, in the end, we are increasing the threshold to which the estimated message innovation rate is compared, and, in practice, also the normal users are expected to exhibit a certain degree of dependence.
%When the BIC fails, the algorithm is expected to fail. Note that it is no longer possible to predict the algorithm behavior, not only as regards the normal users, but even for what concerns the bots. As a matter of fact, the two quantities $\eta_{\textnormal{bot}}$ and $\eta_{\textnormal{nor}}$ are strictly interlaced, since, due to the sequential nature of the algorithm, when the BIC is failing, a certain kind of error (i.e., including a normal user into the estimated bot) might affect the subsequent choices in a rather subtle and unpredictable manner.
%}

The analysis summarized in Fig.~\ref{fig:fig2} reveals that the choice of the threshold is not critical, and the algorithm offers excellent performance for a relatively large range of $\epsilon$. Indeed, recall that $\epsilon\in(0,1)$, and that $\epsilon$ must be ``small'', so that $\epsilon=0.05$ up to $0.2$ can be definitely considered a ``large'', flexible range.

\begin{figure}[t]
\centerline{\includegraphics[width=.45\textheight]{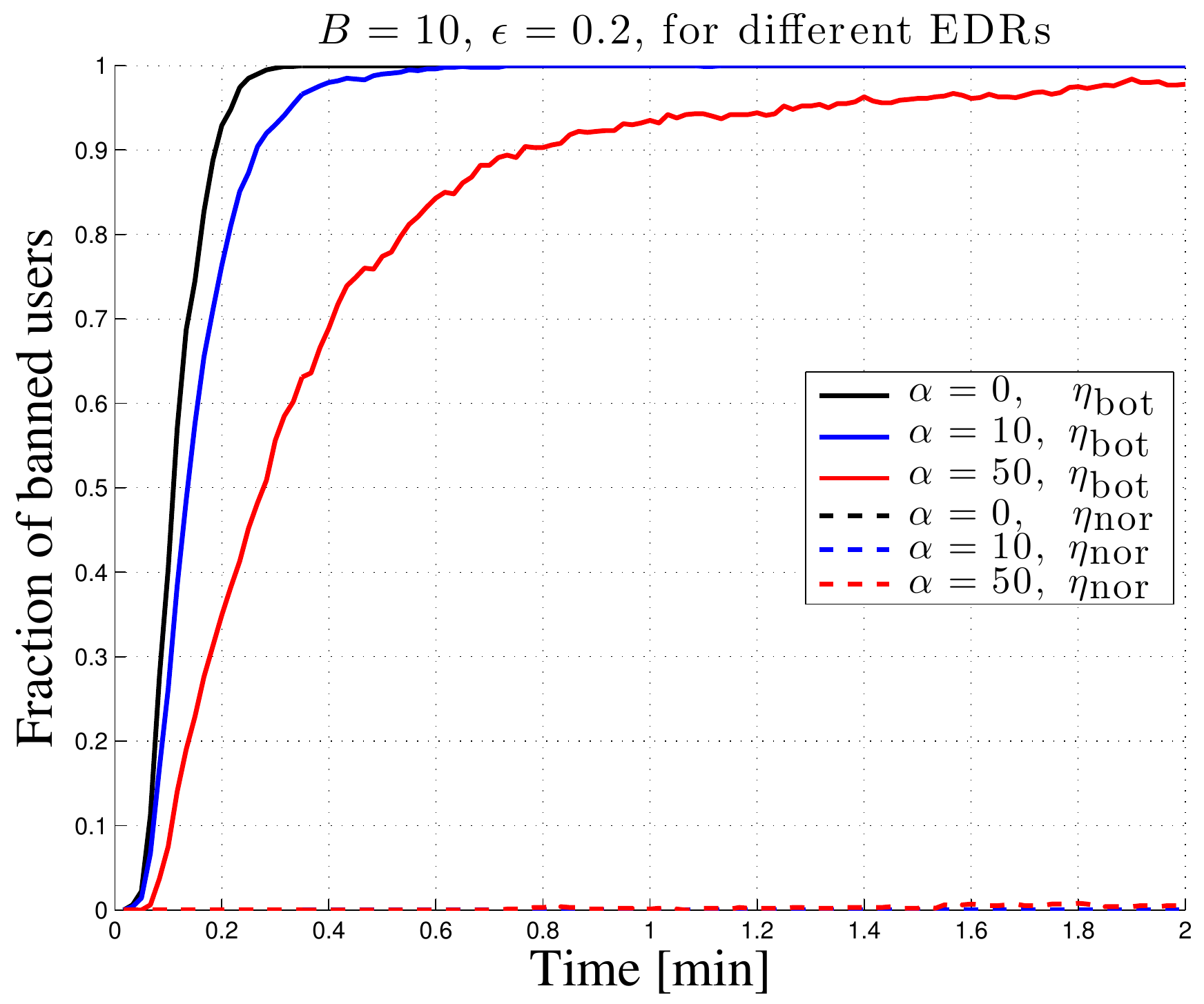}}
\caption{Fraction of banned users as a function of time, for different values of the EDR $\alpha$. 
The monitored network is composed of $10$ normal users, and contains $B=10$ bots. 
Solid curves refer to correctly banned bots, whereas dashed curves refer to erroneously banned normal users.
The depicted curves are computed over $100$ independent Monte Carlo trials. Per each trial, $2$-min chunks of each user are randomly selected among the available chunks.}
\label{fig:fig4}
\end{figure}

In Fig.~\ref{fig:fig4}, the different curves refer to three EDR values (which, we recall, is {\em not known} to the algorithm). The threshold parameter $\epsilon$ was set to $0.2$.
Let us start by examining the behavior of $\eta_{\textnormal{nor}}$.
We see that, irrespectively of the EDR value, $\eta_{\textnormal{nor}}$ stays approximately constant at $0$, which matches our previous evidences and observations. 

Let us switch to the analysis of $\eta_{\textnormal{bot}}$. 
The lowermost curve corresponds to the highest EDR value considered in the figure, namely, to $\alpha=50$. Compared to what we have observed in the real-network traces, such an EDR is a kind of relatively high value. 
We see that the average percentage of correctly identified bots is relatively large  ($>80\%$), even at the beginning of the monitoring activity.  Then, the estimated $\eta_{\textnormal{bot}}$ increases, approaching unity as time elapses, {\em in perfect accordance with the theoretical results of Theorem~$2$}.
%This fact mainly {\em corroborates the BIC}. 
%As already observed in the previous sections, the BIC is neither an asymptotic, nor an average condition. Accordingly, the fact that the estimated $\eta_{\textnormal{nor}}$ is zero might come with no surprise, and seems to confirm markedly the validity of the BIA.

Let us now move toward examining the incidence of the EDR on the algorithm performance. 
We see that the curves corresponding to $\eta_{\textnormal{bot}}$ move upward as $\alpha$ decreases. 
This sounds perfectly reasonable, since $\alpha$ quantifies the learning ability (i.e., the power) of the botnet. 
On the other hand, for each value of $\alpha$, the performance must eventually reach the limiting value of unity after a sufficiently long time.
In particular, the uppermost curve corresponds to the degenerate case $\alpha=0$, namely, to the classical and well-documented case where the botnet uses repeatedly the same patterns. 
As such, the case $\alpha=0$ could be addressed by other (simpler) tools, since a normal user will seldom feature such a small innovation rate in practice. 
In summary, the above analysis emphasizes that the performance decreases with the botnet learning ability $\alpha$.

\begin{figure}[t]
\centerline{\includegraphics[width=.45\textheight]{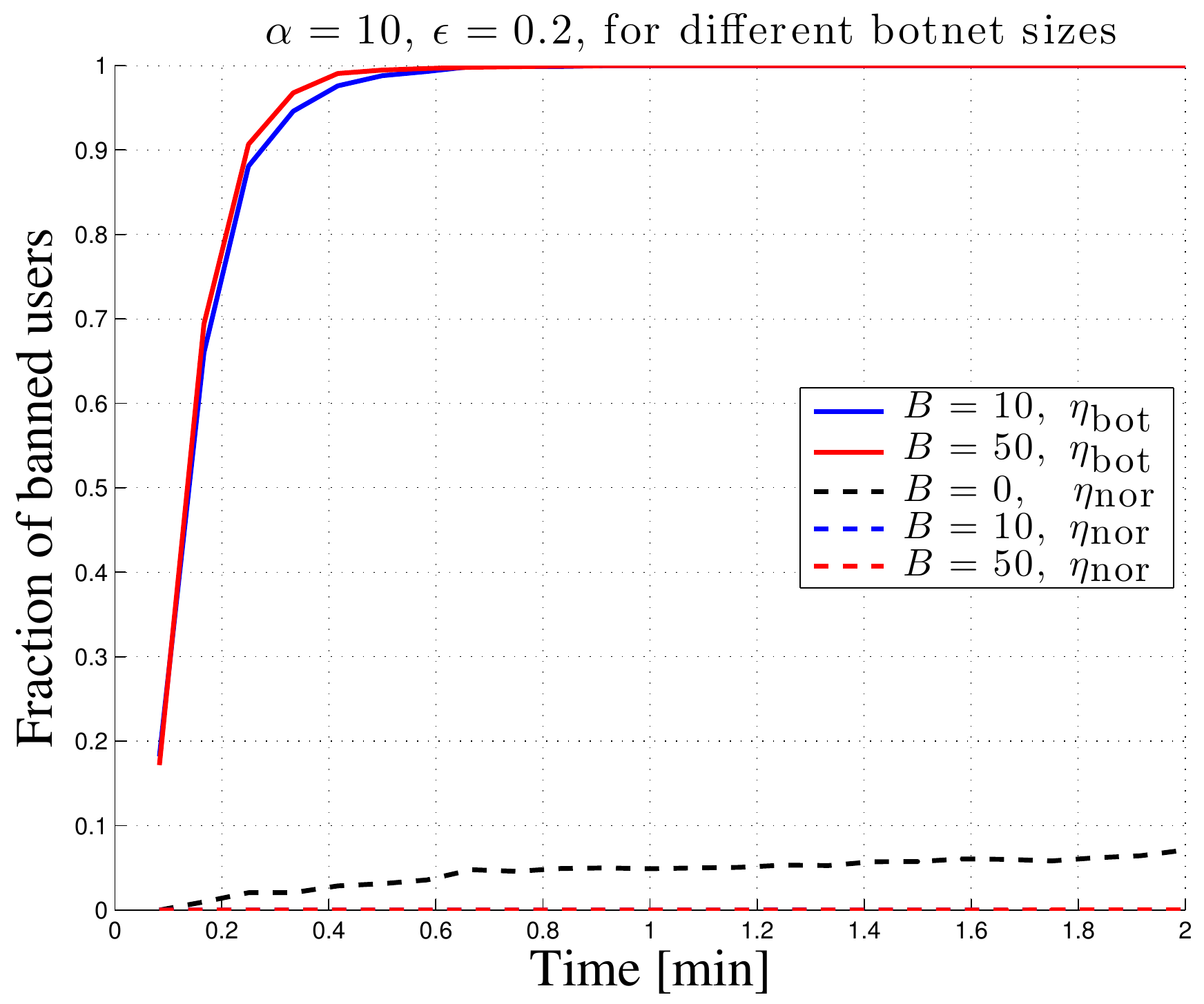}}
\caption{Fraction of banned users as a function of time, for different botnet sizes $B$. 
The monitored network is composed of $50$ normal users (``multiplied'' scenario --- see main text). 
%The normal users have been fictitiously multiplied by treating distinct chunks of the same user as distinct users --- see main text. 
Solid curves refer to correctly banned bots, whereas dashed curves refer to erroneously banned normal users.
The depicted curves are computed over $100$ independent Monte Carlo trials. Per each trial, $2$-min chunks of each user are randomly selected among the available chunks.
}
\label{fig:fig5}
\end{figure}

As last case, we consider the aforementioned scenario where the normal users are fictitiously multiplied by treating distinct chunks of the same user as distinct users. In Fig.~\ref{fig:fig5} we consider a network made of $50$ normal users, for three cases, namely, $B=0$, $B=10$ and $B=50$ bots. The relevant EDR was set to $\alpha=10$, while the threshold parameter was set to the intermediate value $\epsilon=0.2$. In agreement with our previous evidences, even for such larger network, we see that $\eta_{\textnormal{nor}}$ is very small when $B=0$, and is in practice zero for $B=10, 50$.  

We see that $\eta_{\textnormal{bot}}$ increases with $B$. 
However, increasing $B$ is expected to augment the botnet ``visibility'', but also the number of mistakes the algorithm can commit. Thus, the dependence of the performance upon $B$ is not obvious. 
In fact, other evidences collected during our experimental campaign (not reported for space constraints), suggest that $\eta_{\textnormal{bot}}$ is not necessarily monotonically increasing with $B$.

Finally, we stress that simulations were carried for networks up to $100$ nodes. 
The algorithm was able to guarantee the real-time requirement, yet with a standard laptop, with no careful managing of memory and computational burden, and no code optimization, such issues being beyond the scope of the work. 

\begin{appendices}

\section{}
\label{app:A}
In the following, the symbol $o(g_n)$ will denote a function such that $o(g_n)/g_n\rightarrow 0$ as $n\rightarrow \infty$.
Also, when convenient for notational reasons, the expectation of $X$ is denoted by $\bar X$. 

{\bf \textsc{Proposition 1} (Useful recursion).}
{\em Let $a, c>0$, $b\in\mathbb{R}$, $n\in\mathbb{N}$, $\eta_n=1-1/(c+a n)$, and $f_n=\eta_n \, f_{n-1}+b$.
%\label{eq:recursion}
We have: 
\beq
f_n=f_0 \prod_{\ell=1}^n \eta_\ell
+
b\left(
1+\sum_{k=2}^n \prod_{\ell=k}^n \eta_\ell
\right),
\label{eq:trivialrecurs}
\eeq
or:
\beq
f_n=f_0\prod_{\ell=1}^n \eta_\ell
+
\frac{a b}{1+a}
\left[
n+\left(1+\frac c a\right)
\left(1-
\prod_{\ell=1}^n \eta_\ell
\right)
\right],
\label{eq:explicit}
\eeq
and the following limit holds:
\beq
\boxed{
\lim_{n\rightarrow\infty} \frac{f_n}{n}=\frac{a b}{1+a}
}
\label{eq:limit}
\eeq
}
{\em Proof.} First, observe that:  
\beq
f_1=f_0 \eta_1+b,\qquad f_2=f_0 \eta_1\eta_2+b(1+\eta_2),\dots
\label{eq:trivialrecurs2}
\eeq
which yields~(\ref{eq:trivialrecurs}) by recursion. 
Let now $\hat f_n$ denote the RHS in~(\ref{eq:trivialrecurs2}). 
By the induction principle, the claim in~(\ref{eq:trivialrecurs2}) will be proved if we show that $f_1=\hat f_1$, and that 
\beq
\boxed{
f_n=\hat f_n \Rightarrow f_{n+1}=\hat f_{n+1}
}
\eeq
Making explicit the definition of $\eta_1$ where needed, we have: 
\beq
\hat f_1=\eta_1 f_0 + \frac{a b}{1+a}(1+1/a)=\eta_1 f_0 + b=f_1.
\eeq
Assuming now that $f_n=\hat f_n$, we can write
\beqa
\lefteqn{
f_{n+1}=\eta_{n+1}\,\hat f_n+b
=
f_0\prod_{\ell=1}^{n+1} \eta_\ell
-\frac{a b}{1+a}
\left(
1+\frac c a
\right)
\prod_{\ell=1}^{n+1} \eta_\ell
}\nonumber\\
&+&
\underbrace{
\frac{a b}{1+a}
\left(
n+1+\frac c a
\right)
\left(
1-\frac{1}{c+a(n+1)}
\right)
+b
}_{=
\frac{a b}{1+a}\left[
n+1+\left(1+\frac c a\right)
\right]
}
=\hat f_{n+1}.\nonumber\\
\eeqa
Finally, the claim in~(\ref{eq:limit}) follows by observing that the term $\prod_{\ell=1}^n \eta_\ell$ in~(\ref{eq:explicit}), vanishes as $n\rightarrow\infty$.~$\hfill\blacksquare$

{\bf \textsc{Corollary 1} (Small perturbations).}
{\em Let $a, b>0$, $n\in\mathbb{N}$, and let $f_n$ be a nonnegative sequence such that:
\beq
f_n\leq f_{n-1}\left(
1-\frac{1}{a n+o(n)}
\right)+b+o(1).
\label{eq:uppineq}
\eeq
Then:
\beq
\limsup_{n\rightarrow\infty} \frac{f_n}{n}\leq\frac{a b}{1+a}.
\label{eq:claimupper}
\eeq
If the inequality in~(\ref{eq:uppineq}) is reversed, the constant $b$ can be relaxed to be an arbitrary real number, and:
\beq
\liminf_{n\rightarrow\infty} \frac{f_n}{n}\geq\frac{a b}{1+a}.
\label{eq:claimlower}
\eeq
}
{\em Proof.} Clearly, it suffices to prove~(\ref{eq:claimupper}). In the following, $\epsilon>0$ is an arbitrarily small constant. For $n$ large enough, and for all $c\in\mathbb{R}$, we have:
\beq
0<1-\left(\frac{1}{a n+o(n)}\right)
\leq1-\frac{1-\epsilon}{c+a n}.
\eeq
Moreover, we have $b+o(1)\leq b+\epsilon$. Since $f_n$ is nonnegative by assumption, a certain $n_0$ exists, such that, for all $n>n_0$:
\beq
f_n\leq f_{n-1}\left(1-\frac{1-\epsilon}{c+a n}\right)+b+\epsilon.
\label{eq:ineqchain}
\eeq
Introducing, for $m=1,2,\dots$, the definition 
\beq
\eta_m=1-\frac{1-\epsilon}{c+a (n_0+m)}=1-\frac{1}{\frac{c+a n_0}{1-\epsilon}+\frac{a}{1-\epsilon}\,m}.
\label{eq:etafictdef}
\eeq
from~(\ref{eq:ineqchain}) we get, by recursion:
\beq
f_{n_0+m}\leq f_{n_0} \prod_{\ell=1}^m \eta_\ell +
(b+\epsilon)\left(
1+\sum_{k=2}^m \prod_{\ell=k}^m \eta_\ell
\right).
\label{eq:trivialrecurs2}
\eeq
In view of~(\ref{eq:etafictdef}), Proposition 1 allows to conclude that:
\beq
\limsup_{n\rightarrow\infty} \frac{f_n}{n}
\leq
\frac{\frac{a}{1-\epsilon} (b+\epsilon)}{1+\frac{a}{1-\epsilon}},
\eeq
and, hence, the claim in~(\ref{eq:claimupper}) follows from arbitrariness of $\epsilon$.~$\hfill\blacksquare$

{\em Proof of Theorem~$1$}. 
First, we prove the claim for the synchronous scheduling, where all bots transmit regularly at intervals of constant duration $\tau=1/\lambda$. 
Accordingly, we consider a {\em slotted} system with discrete time index $n\geq 0$, and introduce the quantities:
\beq
\mathscr{D}_n\dfz\mathscr{D}_{\mathcal{B}}(n\tau),~M_n\dfz |\mathscr{D}_n|,
~
\mathscr{E}_n\dfz\mathscr{E}(n\tau),~e_n\dfz |\mathscr{E}_n|,
\label{eq:slotted12}
\eeq
where we further observe that:
\beq
\lim_{n\rightarrow\infty}\frac{e_n}{n\tau}=\alpha\Rightarrow
e_n=\alpha\tau\,n + o(n).
\label{eq:leadingordereq}
\eeq
Now, for the synchronous case, it suffices to show that: 
\beq
\frac{M_n}{n\tau}\stackrel{\textnormal{p}}{\longrightarrow} \frac{\alpha B \lambda}{\alpha+B \lambda}
\Leftrightarrow
\frac{M_n}{n}\stackrel{\textnormal{p}}{\longrightarrow} \frac{\alpha\tau\,B}{\alpha\tau+B}\dfz \rho,
\label{eq:equivrhoconverg}
\eeq
where $B$ is the cardinality of subnet $\mathcal{B}$. Observe preliminarily that, by the orthogonality principle, we can write:
\beq
\E\left[\left(\frac{M_n}{n}-\rho\right)^2\right]=
\E\left[\left(\frac{M_n-\bar M_n}{n}\right)^2\right]+
\left(\frac{\bar M_n}{n}-\rho\right)^2,
\label{eq:firstMSequat}
\eeq
%Accordingly, we start by showing that
%\beq
%\lim_{n\rightarrow\infty}
%\frac{\bar M_n}{n}=\rho.
%\eeq
%We have, for $n=1,2,\dots$:
%\beq
%\boxed{
%M_n=M_{n-1}+X_n
%}
%\label{eq:recursively}
%\eeq
and, since mean-square convergence implies convergence in probability~\cite{shao}, it suffices to show that, as $n\rightarrow\infty$, both terms on the RHS in~(\ref{eq:firstMSequat}) vanish.\footnote{In fact, we prove a stronger result in terms of {\em mean-square} convergence.} 
We start by showing that $\bar M_n/n\rightarrow \rho$. At time $n$, the probability that $k$ bots out of $B$ pick a message outside $\mathscr{D}_{n-1}$ is (conditionally on $M_{n-1}$):
\beq
{B\choose k} 
\left(
1-\frac{M_{n-1}}{e_n}
\right)^k
\left(
\frac{M_{n-1}}{e_n}
\right)^{B-k}.
\label{eq:binoprob}
\eeq
Let us introduce the binomial random variable $\hat X_n$, with probability mass function given by~(\ref{eq:binoprob}), whose (conditional) expectation and variance are:
\beq
\E[\hat X_n|M_{n-1}]=B\left(
1-\frac{M_{n-1}}{e_n}
\right),
\label{eq:binomean}
\eeq
and
\beq
\VAR[\hat X_n|M_{n-1}]=B\left(
1-\frac{M_{n-1}}{e_n}
\right)
\frac{M_{n-1}}{e_n}.
\label{eq:binovar}
\eeq
In order to build $\mathscr{D}_n$, we must select all the {\em distinct} messages among the $k$ available ones. 
Ignoring repetitions, we can write:
\beq
M_n\leq M_{n-1}+\hat X_n,
\label{eq:boundontherealizations}
\eeq
and, taking expectations:
\beq
\bar M_n
\leq
\bar M_{n-1}\left(1-\frac{1}{\alpha\tau n/B+o(n)}\right)+B,
\label{eq:preupperbound}
\eeq
having used~(\ref{eq:binomean}) and the expression of $e_n$ appearing on the RHS in~(\ref{eq:leadingordereq}). Direct application of Corollary 1 now yields:
\beq
\limsup_{n\rightarrow\infty} \frac{\bar M_n}{n}\leq\frac{\alpha\tau\,B}{\alpha\tau+B}.
\label{eq:upperbound}
\eeq
Let us now prove the above (reversed) inequality for the $\liminf$. To this aim, we split $\mathscr{E}_n$ into $C$ non-overlapping cells:
\beq
\mathscr{E}_n=\bigcup_{c=1}^C \mathscr{E}_{c,n},\quad
\left\lfloor\frac{|\mathscr{E}_{n}|}{C}\right\rfloor\leq
|\mathscr{E}_{c,n}|\leq
\left\lfloor\frac{|\mathscr{E}_{n}|}{C}\right\rfloor+1,
\label{eq:edictbounds}
\eeq 
where $C$ is an arbitrary integer. 
Since we focus on the regime where $n\rightarrow\infty$, it can be safely assumed that the initial number of words in the emulation dictionary obeys: $e_0\geq C$. Let now:
\beq
\mathscr{D}_n=\bigcup_{c=1}^C \mathscr{D}_{c,n},\quad
M_{c,n}\dfz|\mathscr{D}_{c,n}|,\quad
M_n=\sum_{c=1}^C M_{c,n},
\eeq
and the events, for $j=1,2,\dots,B$, and $c=1,2,\dots,C$:
\beq
{\cal A}_{j,c}\dfz\{\textnormal{bot $j$ picks a message belonging to $\mathscr{E}_{c,n}\setminus \mathscr{D}_{c,n-1}$}\}.
\eeq
Then we have, for any $j$:
\beq
\P[{\cal A}_{j,c}|M_{c,n-1}]=\frac{|\mathscr{E}_{c,n}|-M_{c,n-1}}{|\mathscr{E}_{n}|}\dfz p_{c,n},
\eeq
with the dependence of $p_{c,n}$ upon $M_{c,n-1}$ being suppressed for ease of notation.
From~(\ref{eq:edictbounds}), we have:
\beq
\frac 1 C-\frac 1 e_n -\frac{M_{c,n-1}}{e_n}
\leq p_{c,n}\leq 
\frac 1 C+\frac 1 e_n -\frac{M_{c,n-1}}{e_n}.
\label{eq:pcnbounds}
\eeq
Now, $M_{c,n-1}$ increases by {\em at least} $1$ whenever {\em at least} one bot picks a new message belonging to the $c$-th cell. This implies:
\beq
\E[M_{c,n}|M_{c,n-1}]
\geq
M_{c,n-1} + B p_{c,n} - (B p_{c,n})^2,
\label{eq:prelowerbound}
\eeq
where we used the inequality $(1-p)^B\leq 1-B p+(B p)^2$. 
On the other hand, for large $n$ and small $\epsilon>0$, from~(\ref{eq:pcnbounds}), we get 
$p_{c,n}^2\leq (1/C+1/e_n)^2\leq C^{-2}+\epsilon$, and, hence, from~(\ref{eq:prelowerbound}):
\beq
\E[M_{c,n}|M_{c,n-1}]\geq M_{c,n-1}+B p_{c,n} -\left(\frac{B}{C}\right)^2-\epsilon^\prime,
\eeq
for a certain small $\epsilon^\prime$.
Conversely, using the lower bound in~(\ref{eq:pcnbounds}), and averaging over $M_{c,n-1}$, for large $n$ we get:
\beq
\bar M_{c,n}\geq\bar M_{c,n-1}\left(1-\frac{B}{e_n}\right)+\frac{B}{C}\left(1-\frac{B}{C}\right)-\epsilon^{\prime\prime},
\eeq
for a certain small $\epsilon^{\prime\prime}$.
Summing over $c$, we get:
\beqa
\bar M_n&\geq&
\bar M_{n-1}\left(1-\frac{B}{e_n}\right) + \underbrace{B\left(1-\frac{B}{C}\right)-C\,\epsilon^{\prime\prime}}_{b}\nonumber\\
&=&
\bar M_{n-1}\left(1-\frac{1}{\alpha\tau/B+o(n)}\right)+b,
\eeqa
having used $e_n$ in~(\ref{eq:leadingordereq}). Invoking now Corollary 1, we obtain:
\beq
\liminf_{n\rightarrow\infty}\frac{\bar M_n}{n}\geq 
\frac{\alpha\tau\,b}{\alpha\tau+B}\geq 
\frac{\alpha\tau\,B}{\alpha\tau+B},
\label{eq:veryintermediate}
\eeq
where the latter inequality follows from the definition of $b$, since $C$ and $\epsilon$ are arbitrary.
%\beq
%\liminf_{n\rightarrow\infty}\frac{\bar M_n}{n}\geq 
%\frac{\alpha\tau\,B}{\alpha\tau+B},
%\eeq
Equation~(\ref{eq:veryintermediate}), along with~(\ref{eq:upperbound}), yields that the second term on the RHS in~(\ref{eq:firstMSequat}) vanishes. 
Let us switch to the first term in~(\ref{eq:firstMSequat}). 
In view of the ascertained convergence of expectations, the variance will be proved to vanish if we show that: 
$\E[M_n^2]/n^2\rightarrow\rho^2$. Now, in the light of~(\ref{eq:boundontherealizations}), we can write: 
$\E[M_n^2|M_{n-1}]\leq
M^2_{n-1}+\E[\hat X_n^2|M_{n-1}]
+2\,M_{n-1}\E[\hat X_n|M_{n-1}]$, 
which, using~(\ref{eq:binomean}) and~(\ref{eq:binovar}), yields:
\beqa
v_n&\leq& v_{n-1}\frac{n-1}{n}\left[
1-\frac{2 B}{e_n}+\frac{B(B-1)}{e_n^2}\right]
\nonumber\\
&+&B\frac{\bar M_{n-1}}{n}\left(
2-\frac{2B-1}{e_n}
\right)+\frac{B^2}{n},
\label{eq:vninequal}
\eeqa
having also introduced the definition $v_n\dfz\E[M_n^2]/n$.
%%%%%
%%%%%
\begin{figure*}[!t]
\normalsize
% Store the current equation number.
\setcounter{mytempeqncnt}{\value{equation}}
% Set the equation number to one less than the one
% desired for the first equation here.
% The value here will have to changed if equations
% are added or removed prior to the place these
% equations are referenced in the main text.
\setcounter{equation}{73}
\beq
\hat{\mathcal{B}}_1=\{1\},
\quad
\mathcal{E}_2=\left\{\hat\rho_{\hat{\mathcal{B}}_1\cup\{2\}}<\gamma(\hat{\mathcal{B}}_1,\{2\})\right\},
\quad
\mathcal{E}_3=\left\{\hat\rho_{\hat{\mathcal{B}}_2\cup\{3\}}<\gamma(\hat{\mathcal{B}}_2,\{3\})\right\},
~
\dots
~
\mathcal{E}_{B}=\left\{\hat\rho_{\hat{\mathcal{B}}_{B-1}\cup\{B\}}<\gamma(\hat{\mathcal{B}}_{B-1},\{B\})\right\},
\nonumber
\eeq
\beq
\mathcal{E}_{B+1}=\left\{\hat\rho_{\hat{\mathcal{B}}_B\cap\{B+1\}}
\geq\gamma(\hat{\mathcal{B}}_B,\{B+1\})
\right\},
~
\dots
~
\mathcal{E}_{N}=\left\{\hat\rho_{\hat{\mathcal{B}}_B\cap\{N\}}\geq\gamma(\hat{\mathcal{B}}_B,\{N\})\right\}.
\label{eq:superequat}
\eeq
% Restore the current equation number.
\setcounter{equation}{\value{mytempeqncnt}}
\hrulefill
%\vspace*{4pt}
\end{figure*}
%%%%%%
%%%%%%
Now, the first term appearing on the RHS can be represented as
\beq
v_{n-1}\left(1-\frac{1}{\frac{\alpha\tau}{\alpha\tau+2 B} n + o(n)}\right).
\eeq
Likewise, the second term appearing on the RHS in~(\ref{eq:vninequal}) can be written as $2B\rho+o(1)$.
Applying Corollary 1, we get:
\beq
\limsup_{n\rightarrow\infty}\frac{\E[M^2_n]}{n^2}
=
\limsup_{n\rightarrow\infty}\frac{v_n}{n}
\leq
2 B\rho\,\frac{\frac{\alpha\tau}{\alpha\tau+2 B}}{1+\frac{\alpha\tau}{\alpha\tau+2 B}}
=\rho^2.
%\nonumber\\
%&=&2 B\,\frac{\alpha\tau\,B}{\alpha\tau+B}\,\frac{\alpha\tau}{2 (\alpha\tau+B)}=
\label{eq:limsupsup}
\eeq
Now, subadditivity of limit superior implies:
%\footnote{The limit superior of the pertinent summands is finite.}
\beqa
\lefteqn{
\limsup_{n\rightarrow\infty}
\E\left[\left(\frac{M_n-\bar M_n}{n}\right)^2\right]
}\nonumber\\
&\leq&
\limsup_{n\rightarrow\infty}\frac{\E[M^2_n]}{n^2}+
\limsup_{n\rightarrow\infty}\left(-\frac{\bar M^2_n}{n^2}\right)\leq 0,
\eeqa
with the latter inequality coming from~(\ref{eq:limsupsup}), and from $\bar M_n/n\rightarrow \rho$. 
The claim for the synchronous case is so proved.

With regard to the Poisson case, we consider again the slotted system in~(\ref{eq:slotted12}), but for the fact that $\tau$ is now an arbitrarily {\em small} interval. 
The number of transmission attempts in a single slot, $A$, is now a Poisson random variable with expectation $\bar{A}=\sum_{u\in\mathcal{B}} \lambda_u \tau=\lambda_{\mathcal{B}}\tau$. 
%$\P[A=m]=(\bar{A}^m/m!)\, e^{-\bar{A}}$
Since the $A$ transmissions correspond to $A$ independent choices of messages from the emulation dictionary, for small $\tau$ the system behaves as if we had $A$ synchronous bots, where $A$ is now {\em random}. 
Thus, the proof for the Poisson case boils down to modify slightly the previous proof in order to take into account such additional randomness. Specifically, Eq.~(\ref{eq:preupperbound}) should be modified by considering a random number of bots $A$, and then taking expectations, yielding:\footnote{We implicitly use: $i)$ the independence between  scheduling policy and message picking, and $ii)$ the memoryless property of the Poisson process. 
}
$\bar M_n\leq \bar M_{n-1}(
1-\bar{A}/e_n
)+\bar{A}$. 
Likewise, Eq.~(\ref{eq:prelowerbound}) becomes:
$\E[M_{c,n}|M_{c,n-1}]\geq M_{c,n-1}+1-\E[(1-p_{c,n})^{A}|M_{n-1}]$. Since, for the Poisson random variable $A$, it is easy to show that $\E[(1-p)^{A}]=e^{-\bar{A}p}\leq 1-\bar{A}p+(\bar{A}p)^2$, the conclusion in~(\ref{eq:equivrhoconverg}) still holds true, with $B$ simply replaced by $\bar{A}$. Finally, the inequality in~(\ref{eq:vninequal}) becomes:
\beqa
v_n&\leq& v_{n-1}\frac{n-1}{n}\left[
1-\frac{2 \bar{A}}{e_n}+\frac{\overline{A(A-1)}}{e_n^2}\right]
\nonumber\\
&+&\frac{\bar M_{n-1}}{n}\left(
2\bar{A}-\frac{\overline{A(2A-1)}}{e_n}
\right)+\overline{A^2}.
\eeqa
Having shown that all the equations used to prove the pertinent 
convergence hold true with $B$ replaced by $\bar{A}$, we conclude that:
$
\frac{M_n}{n}\stackrel{\textnormal{p}}{\longrightarrow}\frac{\alpha\tau\,\bar{A}}{\alpha\tau+\bar{A}}=
\frac{\alpha\lambda_{\mathcal{B}}}{\alpha+\lambda_{\mathcal{B}}}.
$
$\hfill\blacksquare$

\section{}
\label{app:C}
{\em Proof of Theorem~$2$}. 
Let us focus on a single step of the BotBuster loop, i.e., the algorithm behavior for a fixed $b_0$. 
Consider first the case that $b_0$ is a normal user, and introduce, for $j\in\mathcal{N}\setminus\{b_0\}$, the events: 
\beq
\mathcal{E}_j=\{\hat\rho_{\{b_0\}\cup\{j\}}\geq \gamma(\{b_0\},\{j\})\}.
\eeq
Eq.~(\ref{eq:above}) reveals that, for any $j$, $\P[\mathcal{E}_j]\rightarrow 1$ as $t\rightarrow\infty$. 
But we also have that, for $b_0$ normal,
\beq
\P[\textnormal{inner loop outputs $\hat{\mathcal{B}}=\{b_0\}$}]=
\P[\cap_{j\in\mathcal{N}\setminus\{b_0\}}\mathcal{E}_j]
\rightarrow 1,
\eeq
where the convergence follows by the fact that each of the events has probability converging to one as $t\rightarrow\infty$.

In contrast, if $b_0$ is a bot, we distinguish two cases: $i)$ if $j$ is normal, from~(\ref{eq:above}) we conclude that $\hat\rho_{\{b_0\}\cup\{j\}}\geq \gamma(\{b_0\},\{j\})$ with probability converging to one as $t\rightarrow\infty$, while $ii)$ if $j$ is a bot, from~(\ref{eq:below}) we conclude that $\hat\rho_{\{b_0\}\cup\{j\}}< \gamma(\{b_0\},\{j\})$ with probability converging to one as $t\rightarrow\infty$. 
%%%%%%%%%%%%%%%%%%%%%%%%%
% Account for the double column equations here.
\addtocounter{equation}{1}
%%%%%%%%%%%%%%%%%%%%%%%%%
Assume now, without loss of generality, that the first $B$ users are bots, that $b_0=1$, and that the remaining users are normal.
In~(\ref{eq:superequat}), we introduce the events corresponding to the inner loop over index $j$, as well as the associated botnet estimates at step $j$, denoted by $\hat{\mathcal{B}}_j$. After noticing that, in the definition of these events, the inequality signs in the threshold comparisons are different for $j\leq B$ and for $j>B$, it is seen that 
the event $\hat{\mathcal{B}}=\{1,2,\dots,B\}$ corresponds to the event $\cap_{j=2}^{N}\mathcal{E}_j$.
Since, in view of the above points $i)$ and $ii)$, we have $\P[\mathcal{E}_j]\rightarrow 1$, we conclude that (if $b_0=1$ is a bot):
\beq
\P[\textnormal{inner loop outputs $\hat{\mathcal{B}}=\{1,2,\dots,B\}$}]
=
\P[\cap_{j=2}^{N}\mathcal{E}_j]\rightarrow 1,
\eeq
which implies the validity of~(\ref{eq:etabotnorconv}).~$\hfill\blacksquare$

\end{appendices}


\end{document}